\documentclass[%
 reprint,
superscriptaddress,
showpacs,
 amsmath,amssymb,
 amsfonts, 
 aps,
prapplied,
floatfix,
]{revtex4-1}

\usepackage{graphicx}
\usepackage{float}

\begin{document}

\title{Optoelectronic and stability properties of quasi-2D alkylammonium based perovskites} 

\author{N. Filipoiu}
\thanks{These authors contributed equall to this work.}
\affiliation{Horia Hulubei National Institute for Physics and Nuclear Engineering, 077126 Magurele-Ilfov, Romania}
\affiliation{University of Bucharest, Faculty of Physics, 077125 Magurele-Ilfov, Romania}

\author{Anca G. Mirea}
\thanks{These authors contributed equall to this work.}
\affiliation{National Institute of Materials Physics, Magurele 077125, Ilfov, Romania}

\author{Sarah Derbali}
\affiliation{National Institute of Materials Physics, Magurele 077125, Ilfov, Romania}

\author{C.-A. Pantis-Simut}
\affiliation{Horia Hulubei National Institute for Physics and Nuclear Engineering, 077126 Magurele-Ilfov, Romania}
\affiliation{University of Bucharest, Faculty of Physics, 077125 Magurele-Ilfov, Romania}
\affiliation{Research Institute of the University of Bucharest (ICUB), 90 Panduri Street, 050663 Bucharest, Romania}

\author{D.-V. Anghel}
\affiliation{Horia Hulubei National Institute for Physics and Nuclear Engineering, 077126 Magurele-Ilfov, Romania}
\affiliation{University of Bucharest, Faculty of Physics, 077125 Magurele-Ilfov, Romania}
\affiliation{Research Institute of the University of Bucharest (ICUB), 90 Panduri Street, 050663 Bucharest, Romania}

\author{A. Manolescu}
\affiliation{Department of Engineering, Reykjavik University, Menntavegur 1, IS-102 Reykjavik, Iceland}

\author{Ioana Pintilie}
\affiliation{National Institute of Materials Physics, Magurele 077125, Ilfov, Romania}

\author{Mihaela Florea}
\email{mihaela.florea@infim.ro}
\affiliation{National Institute of Materials Physics, Magurele 077125, Ilfov, Romania}

\author{G. A. Nemnes}
\email{nemnes@solid.fizica.unibuc.ro}
\affiliation{Horia Hulubei National Institute for Physics and Nuclear Engineering, 077126 Magurele-Ilfov, Romania}
\affiliation{University of Bucharest, Faculty of Physics, 077125 Magurele-Ilfov, Romania}
\affiliation{Research Institute of the University of Bucharest (ICUB), 90 Panduri Street, 050663 Bucharest, Romania}


\begin{abstract}
Electronic and stability properties of quasi-2D alkylammonium perovskites are investigated using density functional theory (DFT) calculations and validated experimentally on selected classes of compounds. Our analysis is focused on perovskite structures of formula (A)$_2$(A$'$)$_{n-1}$Pb$_n$X$_{3n+1}$, with large cations A = butyl-, pentyl-, hexylammonium (BA, PA, HXA), small cations A$'$ = methylammonium, formamidinium, ethylammonium, guanidinium (MA,FA,EA,GA) and halogens X = I, Br, Cl. The role of the halogen ions is outlined for the band structure, stability and defect formation energies. Two opposing trends are found for the absorption efficiency versus stability, the latter being assessed with respect to possible degradation mechanisms. 
Experimental validation is performed on quasi-2D perovskites based on pentylammonium cations, namely: (PA)$_2$PbX$_4$ and (PA)$_2$(MA)Pb$_2$X$_7$, synthesized by antisolvent-assisted vapor crystallization. Structural and optical analysis are inline with the DFT based calculations. In addition, the thermogravimetric analysis shows an enhanced stability of bromide and chloride based compounds, in agreement with the theoretical predictions.
\end{abstract}

\maketitle

\section{Introduction}
\label{intro}

The field of perovskite solar cells (PSCs) has witnessed an impressive development as power conversion efficiencies (PCEs) of 25.7\% \cite{nrel} become comparable with standard silicon based technology at a much lower fabrication cost, taking advantage of chemical synthesis routes. Since the first report of a functional PSC by Miyasaka {\it et al.}, where methylammonium lead triiodide (CH$_3$NH$_3$PbI$_3$ or MAPI) was used, several 3D hybrid perovskites were synthesized by replacing the methylammonium (MA) with other small-size molecules suitable to fit the inorganic lead-halogen cage \cite{CHEN2015355,Park_2021,en14175431}. However, only a limited number of organic molecules or inorganic species are suitable to obtain a 3D structure. In addition, substituting lead by tin or germanium was investigated in order to reduce the toxic potential of the perovskite \cite{doi:10.1021/jacs.7b04219,Azhari_2020}. 

A more substantial way to increase the flexibility of the optoelectronic properties is to use the halogen sequence I, Br, Cl and mixtures thereof, which has a larger impact on the band gap and stability. On the other hand, introducing larger organic cations along with small molecules or completely substituting them brought a much larger diversity of hybrid perovskites structures, usually termed quasi-2D perovskites \cite{doi:10.1021/jacs.8b10851,C9EE03757H,KIM2021100405,D1DT02991F}. 

The quasi-2D perovskites usually exhibit larger band gaps compared to their 3D counterparts, as a consequence of quantum confinement, which leads to a less efficient optical absorption and a smaller PCE \cite{https://doi.org/10.1002/anie.201406466}. Larger band gaps can be used in tandem solar cell configurations and can be further tuned by adjusting the number of quasi-3D layers \cite{doi:10.1021/acs.chemmater.6b00847}. However, this drawback is compensated by the enhanced stability due to the hydrophobic moieties present in the large organic cations \cite{https://doi.org/10.1002/cssc.201802992}. Furthermore, larger cations have smaller diffusivity and are less susceptible to migrate, limiting the ion conduction and the J-V hysteresis typically found in PSCs. 

\begin{figure*}[t]
\begin{flushleft}	
	\hspace*{1.0cm}(a)\hspace*{6cm}(b)\hspace*{6cm}(c)\vspace*{-0.1cm}\\	
\end{flushleft}
 \centering
 \includegraphics[height=6cm]{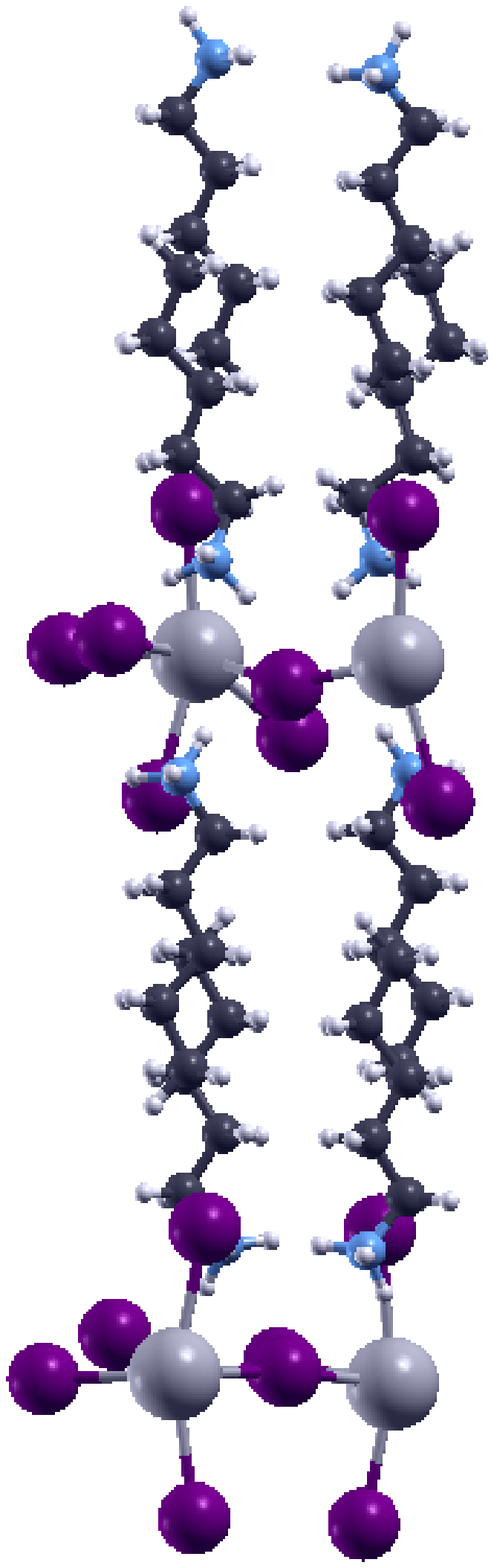}
 \includegraphics[height=6cm]{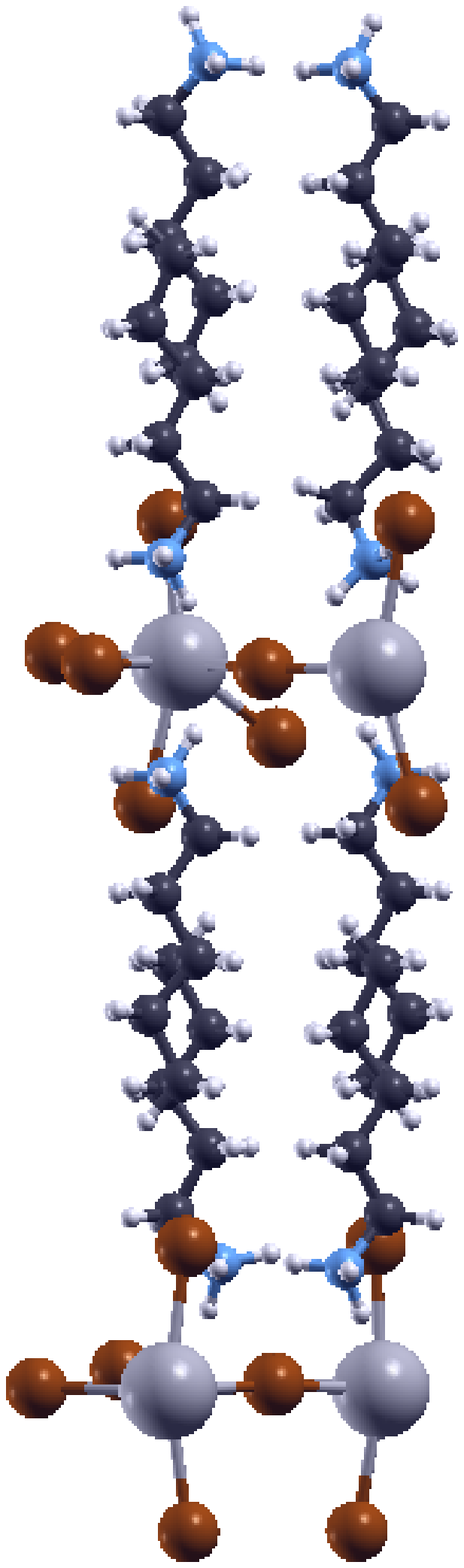}
 \includegraphics[height=6cm]{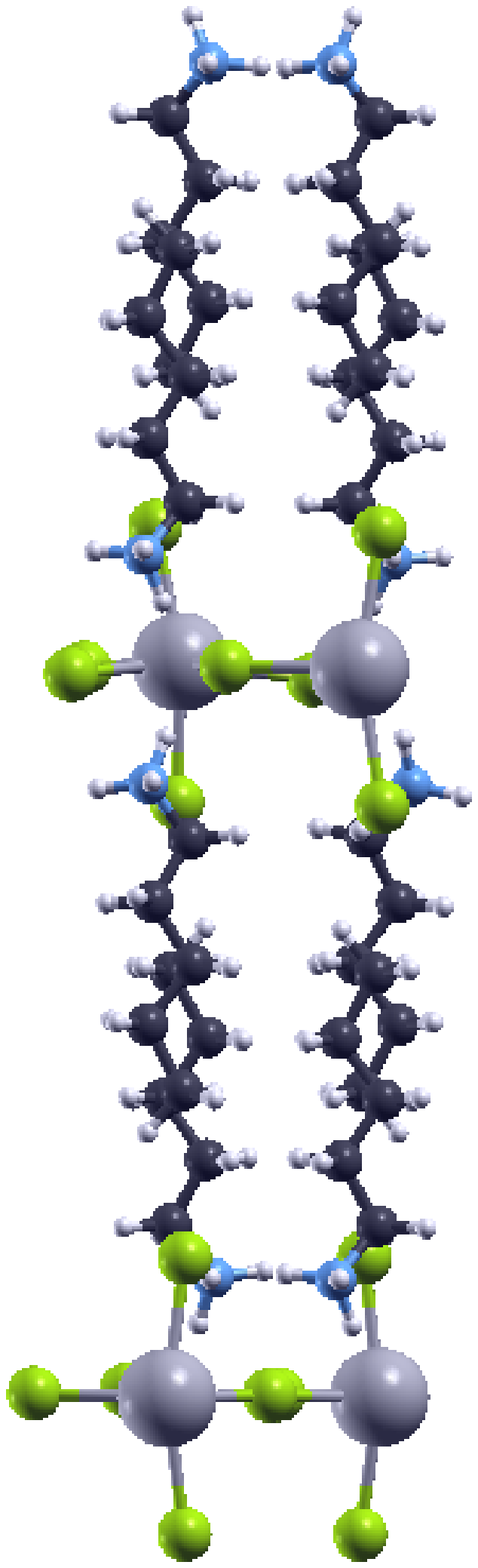}\hspace*{1.5cm}
 \includegraphics[height=6cm]{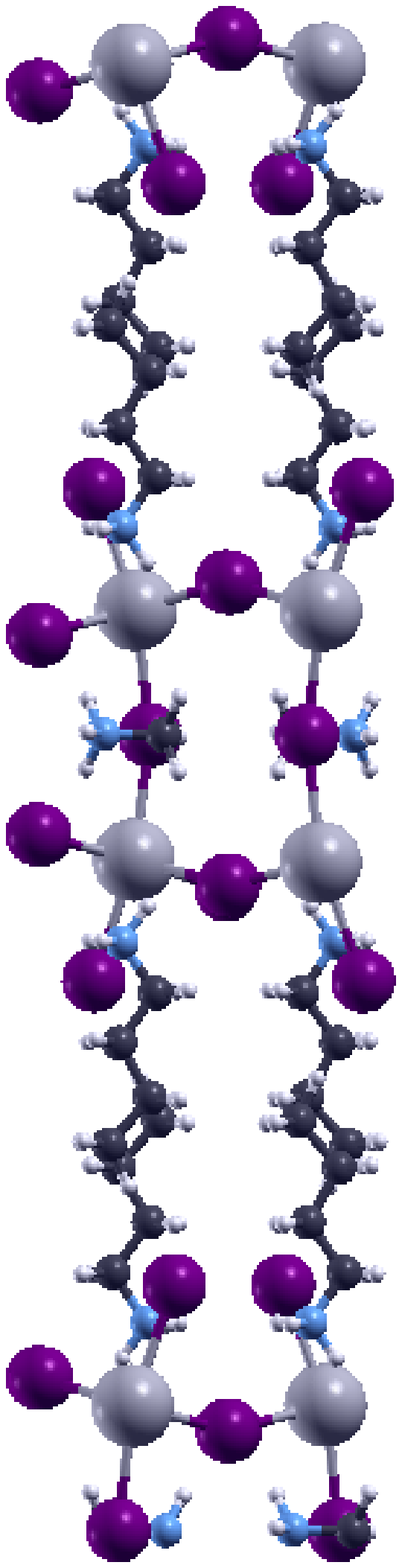}
 \includegraphics[height=6cm]{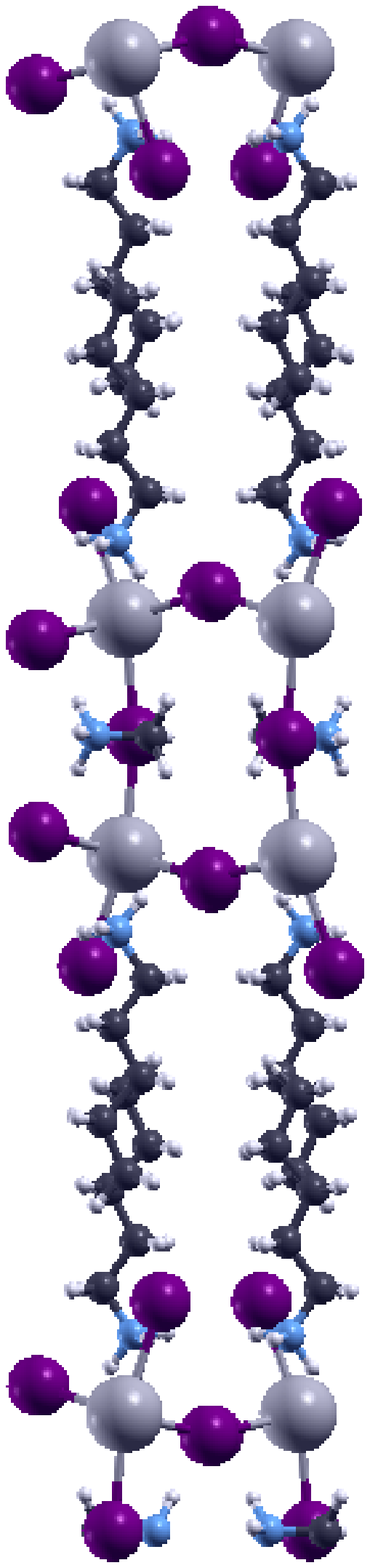}
 \includegraphics[height=6cm]{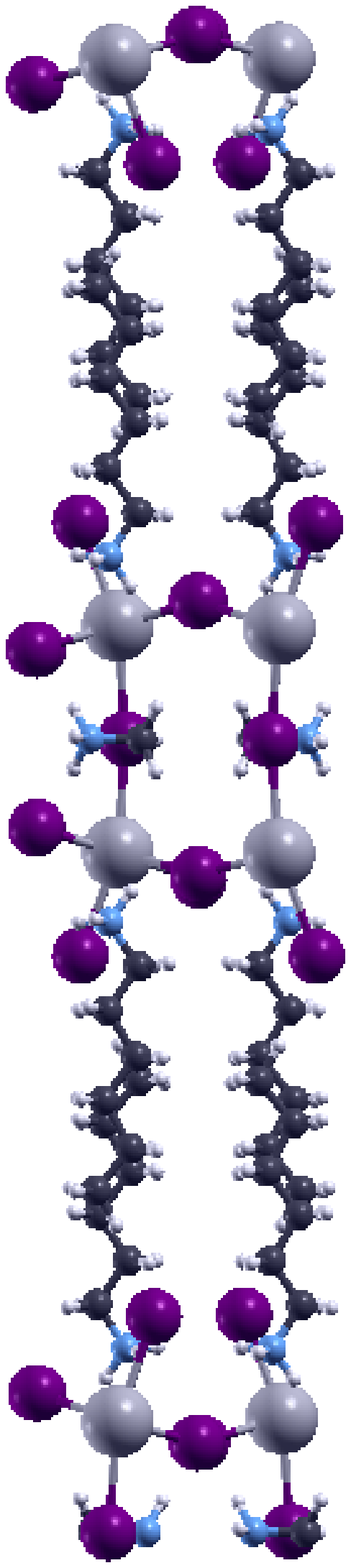}\hspace*{1.5cm}
 \includegraphics[height=6cm]{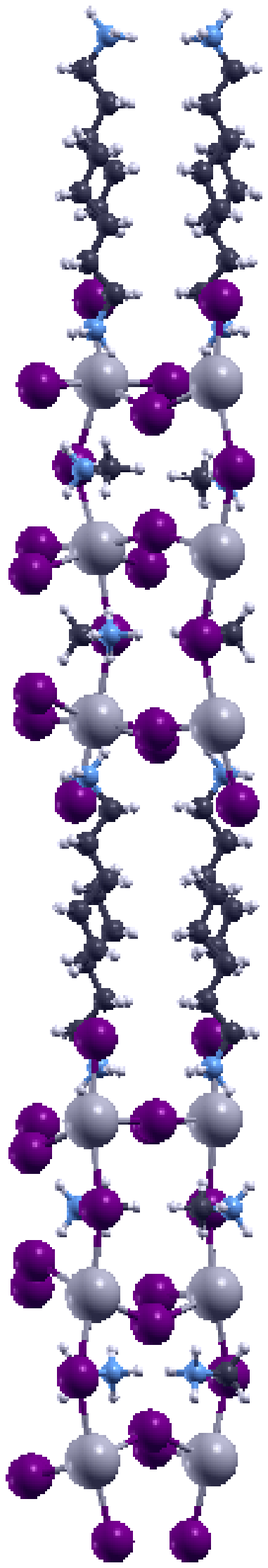}
 \includegraphics[height=6cm]{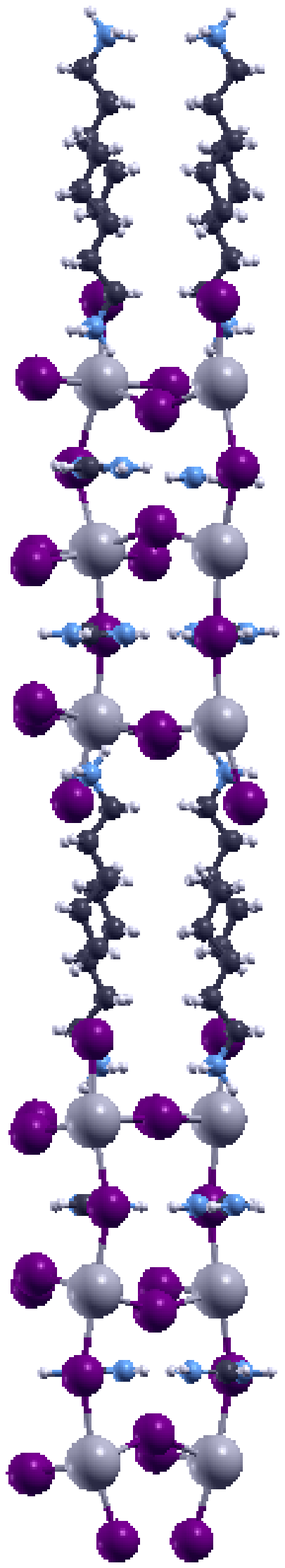}
 \includegraphics[height=6cm]{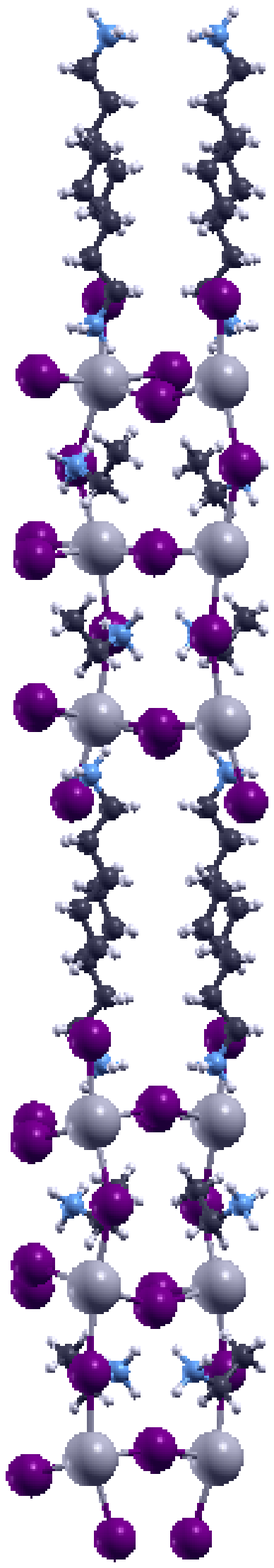}
 \includegraphics[height=6cm]{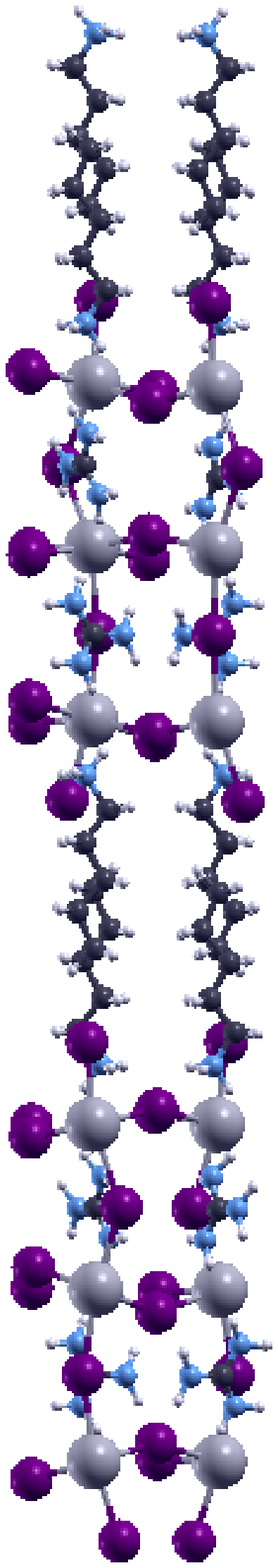}
	\caption{Simulated 2D perovskite structures: (a) (PA)$_2$PbX$_4$, with X = I, Br, Cl; (b) (A)$_2$(MA)Pb$_2$I$_7$, with A = BA, PA, HXA; (c) (PA)$_2$(A')$_2$Pb$_3$I$_{10}$, with A' = MA, FA, EA, GA.}
 \label{structures}
\end{figure*}

From a theoretical perspective, the quasi-2D perovskites have been analyzed using density functional theory (DFT), which allows a computational design for optimal optoelectronic properties\cite{doi:10.1021/acs.jpcc.8b05715,doi:10.1021/acs.jpcc.2c01840}. More and more frequently, the DFT high throughput calculations are optimized using machine learning schemes \cite{Lee2022}. Using first principle methods, the edge states were shown to be stabilized by the internal electric fields induced by polarized  organic cations \cite{doi:10.1021/acs.nanolett.0c03468}. Furthermore, {\it ab initio} calculations also revealed a giant Rashba splitting, which arise from multiple quantum wells with strong spin-orbit coupling \cite{doi:10.1126/sciadv.1700704}, supporting the idea of possible spintronic applications.

Within the multiple choices for the large organic cations, the alkylammonium series offer good prospects for systematic optimizations \cite{billing2007}. A comparative analysis of stability for butylammonium and hexylammonium organolead perovskites has been performed \cite{doi:10.1021/acs.jpcc.0c02822}, evidencing the role of cation length in decreasing the halide mobility. Similarly, the size of diaminoalkane molecules, having two NH2- functional groups, is shown to directly influence the crystallization process \cite{C8CE00999F}. A quasi-2D perovskite containing pentylammonium and guanidinium has been synthesized for a highly sensitive and ultrafast responding photodetector \cite{https://doi.org/10.1002/adom.201900308}.

In this paper, we investigate electronic and stability properties of quasi-2D perovskites based on the alkylammonium series, which includes the butyl-, pentyl- and hexylammonium (BA, PA, HXA) as large cations and methylammonium (MA), formamidinium (FA), ethylammonium (EA), guanidinium (GA) as small cations. This class also allows a systematic analysis concerning the length of the alkyl chain. Furthermore, by changing the halogen, the variations in band gap and stability are determined. The stability is investigated based on the likely degradation mechanisms. In addition, the formation energy of point defects is evaluated for each case, which can bring a better perspective on the ion migration. The theoretical investigations are performed in the framework of {\it ab initio} density functional theory (DFT) calculations, which are supported by experimental synthesis, followed by structural and optical characterization of selected compounds and stability tests. 

The paper is structured as follows. In the next section, the quasi-2D perovskite structures are introduced and the details of DFT based calculations are indicated, which includes the stability assessment and the calculation of defect formation energies. In the following section, the experimental results are described, which include structural and optical characterization, the thermal stability evaluated by thermogravimetric analysis, and the experimental methods.  The conclusions are summarized in the final section.

\section{Computational approach}


We focus our investigation on three sub-classes of the more general (A)$_2$(A$'$)$_{n-1}$Pb$_n$X$_{3n+1}$ Ruddlesden--Popper halide perovskites: (PA)$_2$PbX$_4$ (X = I, Br, Cl), (A)$_2$(MA)Pb$_2$I$_7$ (A = BA, PA, HXA) and (PA)$_2$(A$'$)$_2$Pb$_3$I$_{10}$ (A$'$ = MA, FA, EA, GA). This selection, which belong to the $X_4$, $X_7$ and $X_{10}$ groups, allows a comprehensive perspective concerning the influence of the halogen, and of the large and small cations on the optoelectronic and stability properties. The 2D perovskite compounds crystallize in orthorhombic space groups, which can be either non-centrosymmetric or centrosymmetric. We adopt here the centrosymmetric structures as they resemble the averaged structure of polycrystalline bulk perovskites \cite{doi:10.1021/acs.chemmater.6b00847}, which are depicted in Fig.\ \ref{structures}. The three groups of structures, identified as $X_4$, $X_7$ and $X_{10}$, have the spacegroups $P_{bca}$, $C_{cmm}$ and $A_{cam}$, respectively. The number of atoms per unit cell ranges from 180 ($X_4$) to 276 atoms (PA-MA-X$_{10}$) and 292 atoms (PA-GA-X$_{10}$).  

\subsection{Electronic structure}

We performed {\it ab initio} calculations using the SIESTA package \cite{0953-8984-14-11-302}, which has the advantage of using strictly localized basis sets, allowing a linear scaling of the computational time with the system size. We employed a double-$\zeta$ polarized basis set and local density approximation with an exchange-correlation functional parameterized by Ceperley and Alder \cite{PhysRevLett.45.566}. Even though hybrid functionals and spin-orbit coupling typically provide more accurate band gap energies, there is a fortuitous cancellation that can be exploited, rendering the less expensive functionals to perform reasonably given the relatively large unit cells. The approach makes use of Troullier-Martins norm conserving pseudopotentials \cite{PhysRevB.43.1993} with typical valence configurations. For the k-space sampling a Monkhorst-Pack scheme of $5\times1\times5$ was employed and the real space grid was specified by a mesh cutoff energy of 150 Ry. Structural relaxations are performed until the forces are less than 1 eV/\AA. The density of states is smoothed using a broadening of 0.02 eV.


\begin{figure}[t]
\begin{flushleft}	
  \includegraphics[height=6cm]{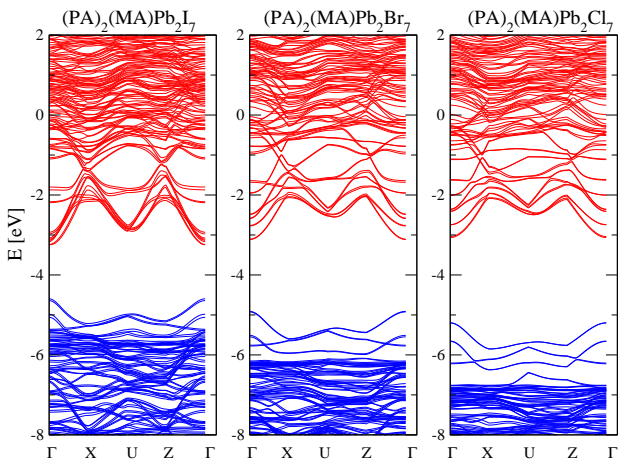} 
\end{flushleft}	
	\caption{Band structures of (PA)$_2$(MA)Pb$_2$X$_7$ quasi-2D perovskites, with X = I, Br, Cl. A direct band gap is evidenced at $\Gamma$ point. The band gap increases as the atomic number of the halogen decreases.}
  \label{PA-MA-X7}
\end{figure}


\begin{figure}[t]
\centering
  \includegraphics[height=6cm]{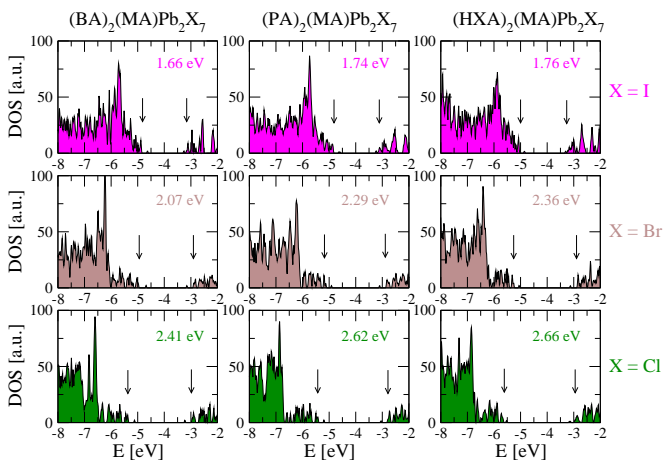}
	\caption{Density of states of (A)$_2$(MA)Pb$_2$X$_7$, with A = BA, PA, HXA and X = I, Br, Cl. The band gaps are strongly influenced by the halogen type, while they are slightly increasing as the size of the A-cation is getting larger.}
  \label{A-MA-X7}
\end{figure}

\begin{figure}[t]
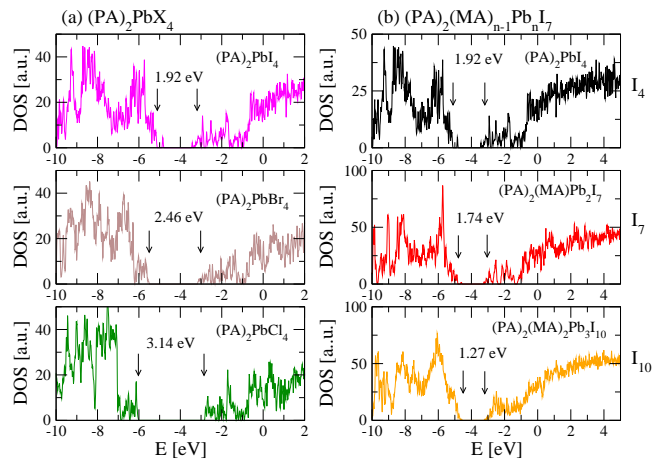

\centering
  \includegraphics[height=6cm]{figure_4a}
  \includegraphics[height=6cm]{figure_4b}
	\caption{Comparing the band gap variation for X$_4$ and $X_7$ quasi-2D perovskite classes: (a) (PA)$_2$PbX$_4$, with X = I, Br, Cl and (b) (PA)$_2$(MA)$_{n-1}$Pb$_n$I$_{3n+1}$, with n = 1, 2, 3. The X$_4$ class shows a similar trend as for the X$_7$ class as the halogen is changed, while increasing $n$ by the amount of MA the band gap is decreasing in the sequence I$_4$ $\searrow$ I$_7$ $\searrow$ I$_{10}$.}
  \label{DOS_PA-X4-X10}
\end{figure}

We start our analysis with (PA)$_2$(MA)Pb$_2$X$_7$ (X = I, Br, Cl) compounds, which are the smallest unit cell structures that contain both large and small cations. The band structures depicted in Fig.~\ref{PA-MA-X7} are similar to previous reports on BA based compounds \cite{doi:10.1021/acs.chemmater.6b00847}. Although the band gaps are typically underestimated by DFT, the trend of increasing gaps in the halogen sequence I$\rightarrow$Br$\rightarrow$Cl is evident. Furthermore, a direct band gap is obtained, which is located at the center of the Brillouin zone.

The gap modifications induced by the size of the large cations have been less analyzed so far. Our choice for the BA, PA, HXA sequence provides the framework for a systematic investigation. Figure\ \ref{A-MA-X7} shows the density of states for the (A)$_2$(MA)Pb$_2$X$_7$ class of compounds (A = BA, PA, HXA ; X = I, Br, Cl). Besides the expected halogen variation, a small increase in the band gap is observed as the size of the large cation is getting larger (BA$\rightarrow$PA$\rightarrow$HXA). This dependence is sub-linear, with a smaller increase in the gaps in the large cation sequence, but systematic for all three halogen types. Larger cations enhance the quantum confinement of the 2D layers containing MA small cations, which, in turn, enhance the band gap with a small amount. 

Removing the small cations, the (A)$_2$(MA)Pb$_2$X$_7$ type structures are transformed into X$_4$ structures of formula (A)$_2$PbX$_4$, which exhibit an even larger quantum confinement. This becomes visible in Fig.\ \ref{DOS_PA-X4-X10}(a), where a similar trend for the same halogen sequence is observed for the (PA)$_2$PbX$_4$, however, with significantly larger gaps compared to the X$_7$ systems. On the other hand, increasing the number of layers X$_4$ $\rightarrow$ X$_7$ $\rightarrow$ X$_{10}$ the quantum confinement is reduced and so is the energy gap. This is illustrated in Fig.\ \ref{DOS_PA-X4-X10}(b) for iodine based perovskites.

\subsection{Stability}

The stability assessment is performed based on the most likely degradation mechanisms. In general, these can be grouped into {\it intrinsic} degradation mechanisms, and {\it extrinsic} ones, similar to reaction routes available for 3D perovskites \cite{acsenergylett.9b01605,C7SE00114B,en14175431}. In the former class, chemically pristine compounds can decompose under thermal and electrical stress, while the latter class belongs to processes where extrinsic elements are involved, like molecular oxygen and water. In the following, we adapt these degradation mechanisms to quasi-2D perovskites. From the former group of intrinsic degradation pathways, we consider the decomposition into a solid precipitate (PbX$_2$) and gaseous phases (organic cations -- A, A$'$ and HX acids):
\begin{eqnarray}
\label{int_degr}
& & (\text{A})_2(\text{A}')_{n-1}\text{Pb}_n\text{X}_{3n+1} \; \rightarrow \nonumber\\
& & 2\tilde{\text{A}} \; + \; (n-1)\tilde{\text{A}'}
+ \; (n+1)\text{HX} \; + \; n\text{PbX}_2 \, .
\end{eqnarray}
Here, $\tilde{\text{A}}$ and $\tilde{\text{A}}'$ represent the deprotonated, electrically neutral molecules corresponding to the large and small cations, respectively. 

As {\it extrinsic} degradation mechanism we investigate the impact of molecular oxygen, using the reaction pathway:
\begin{eqnarray}
\label{ext_degr}
 & & (\text{A})_2(\text{A}')_{n-1}\text{Pb}_n\text{X}_{3n+1} \; + \; \left(\frac{n+1}{4}\right)\text{O}_2 \; \rightarrow \nonumber\\ 
 & & 2\tilde{\text{A}} \; + \; (n-1)\tilde{\text{A}'} \nonumber\\
 & & + \; \left(\frac{n+1}{2}\right)\text{H}_2\text{O} \; + \; \left(\frac{n+1}{2}\right)\text{X}_2 \; + \; n\text{PbX}_2 \, .
\end{eqnarray}
The stability is evaluated based on the endothermic reaction energies.

In addition to perovskite decomposition, a comparative analysis of single vacancy defects for different perovskite compositions can  provide further information regarding ion migration, which is correlated to the perovskite degradation and can be identified in the hysteretic effects in the J-V characteristics \cite{doi:10.1021/acs.jpclett.6b02375}. The formation energy ($E_{\rm f}$) for a single-vacancy (X = I, Br, Cl or Pb) is defined as:
\begin{equation}
\label{Eform}	
	E_{\rm f} = E_{\rm prv}^{\rm def} + E_{\rm sp} - E_{\rm prv}^{\rm ideal} \, ,
\end{equation}	
where $ E_{\rm prv}^{\rm def}$ and $E_{\rm prv}^{\rm ideal}$ are the total energies of the defected and ideal crystals, respectively, and $E_{\rm sp}$ is the total energy of the excluded species, equal to half of the total energy of the X$_2$ molecule or the chemical potential in bulk Pb.

\begin{table}[t]
\small
	\caption{Relative stability of X$_4$, X$_7$ and X$_{10}$ quasi-2D perovskites with respect to the intrinsic degradation mechanism [Eq. (\ref{int_degr})].}
  \label{tbl-int}
(a) Changing the halogen in X$_4$ structures:\\
  \begin{tabular*}{0.48\textwidth}{@{\extracolsep{\fill}}|c|ccc|}
    \hline
    \hline
	  {\bf 2D perovskites (X$_4$)} & {\bf I} & {\bf Br} & {\bf Cl} \\
    \hline
	  (PA)$_2$PbX$_4$ & REF & +5\% & +3\% \\
    \hline
    \hline
  \end{tabular*} \vspace*{0.2cm}\\
	(b) Changing the large cation (A) in $X_7$ structures:\\
  \begin{tabular*}{0.48\textwidth}{@{\extracolsep{\fill}}|c|ccc|}
    \hline
    \hline
	  {\bf 2D perovskites (X$_7$)} & {\bf I} & {\bf Br} & {\bf Cl} \\
    \hline
	  (BA)$_2$(MA)Pb$_2$X$_7$ & REF & +4\% & -3\% \\
    \hline
	  (PA)$_2$(MA)Pb$_2$X$_7$ & +3\% & +8\% & +2\% \\
    \hline
	  (HXA)$_2$(MA)Pb$_2$X$_7$ & +4\% & +7\% & -3\% \\
    \hline	  
    \hline	  
  \end{tabular*} \vspace*{0.2cm}\\
	(c) Changing the small cation (A') in X$_{10}$ structures:\\
  \begin{tabular*}{0.48\textwidth}{@{\extracolsep{\fill}}|c|cccc|}
    \hline
    \hline
	  {\bf 2D perovskites (X$_{10}$)} & {\bf MA} & {\bf FA} & {\bf EA} & {\bf GA} \\
    \hline
	  (PA)$_2$(A$'$)$_2$Pb$_3$I$_{10}$ & REF & +6\% & +1\% & +9\% \\
    \hline
	  \hline
  \end{tabular*}
	
\end{table}
\begin{table}[t]
\small
	\caption{Similar to Table\ \ref{tbl-int}, the relative stability of X$_4$, X$_7$ and X$_{10}$ quasi-2D perovskites with respect to the extrinsic degradation mechanism [Eq. (\ref{ext_degr})], involving molecular O$_2$.}
  \label{tbl-ext}
(a) Changing the halogen in X$_4$ structures:\\
  \begin{tabular*}{0.48\textwidth}{@{\extracolsep{\fill}}|c|ccc|}
    \hline
    \hline
	  {\bf 2D perovskites (X$_4$)} & {\bf I} & {\bf Br} & {\bf Cl} \\
    \hline
	  (PA)$_2$PbX$_4$ & REF & +5\% & +6\% \\
    \hline
    \hline
  \end{tabular*} \vspace*{0.2cm}\\
	(b) Changing the large cation (A) in $X_7$ structures:\\
  \begin{tabular*}{0.48\textwidth}{@{\extracolsep{\fill}}|c|ccc|}
    \hline
    \hline
	  {\bf 2D perovskites (X$_7$)} & {\bf I} & {\bf Br} & {\bf Cl} \\
    \hline
	  (BA)$_2$(MA)Pb$_2$X$_7$ & REF & +5\% & +6\% \\
    \hline
	  (PA)$_2$(MA)Pb$_2$X$_7$ & $\sim$0\% & +5\% & +5\% \\
    \hline
	  (HXA)$_2$(MA)Pb$_2$X$_7$ & +1\% & +7\% & +6\% \\
    \hline	  
    \hline	  
  \end{tabular*} \vspace*{0.2cm}\\
	(c) Changing the small cation (A') in X$_{10}$ structures:\\
  \begin{tabular*}{0.48\textwidth}{@{\extracolsep{\fill}}|c|cccc|}
    \hline
    \hline
	  {\bf 2D perovskites (X$_{10}$)} & {\bf MA} & {\bf FA} & {\bf EA} & {\bf GA} \\
    \hline
	  (PA)$_2$(A$'$)$_2$Pb$_3$I$_{10}$ & REF & +2\% & $\sim$0\% & +3\% \\
    \hline
	  \hline
  \end{tabular*}
	
\end{table}

The stability of the perovskite compounds is assessed with respect to the intrinsic and extrinsic degradation pathways outlined by Eqs.\ (\ref{int_degr}) and (\ref{ext_degr}) on three representative sub-classes: (PA)$_2$PbX$_4$, (A)$_2$(MA)Pb$_2$X$_7$ and (PA)$_2$(A$'$)$_2$Pb$_3$I$_{10}$. The relative stability, measured as endothermic reaction energies, is summarized in Tables\ \ref{tbl-int} and \ref{tbl-ext}, taking one compound as reference from each class, which point out the influence of the halogen, small- and large cations on the degradation processes. For the intrinsic degradation mechanism, replacing I with Br leads to higher stability (4\%--8\%), which is confirmed also in the case of the extrinsic degradation mechanism (5\%--7\%). Replacing iodine with chlorine leads to an overall poorer stability with respect to the intrinsic degradation pathway (-3\%--2\%) for the X$_7$ family, while for the X$_4$ compounds the stability of chlorine based compounds is found in between iodine and bromine based compounds. This is in contrast with the extrinsic degradation, where the use of chlorine is beneficial, the improvement with respect to iodine based systems being similar to the one of bromine based compounds (5\%--6\%). The overall behavior of the stability with respect to the halogen type is, to some degree, similar to one corresponding to 3D perovskites \cite{en14175431}: for the intrinsic mechanism, the stability increases in the sequence I $\nearrow$ Cl $\nearrow$ Br [(PA)$_2$PbX$_4$] and  Cl $\nearrow$ I $\nearrow$ Br [(A)$_2$(MA)Pb$_2$X$_7$], while for the extrinsic mechanism we have I $\nearrow$ Cl $\approx$ Br.

\begin{figure}[t]
\centering
  \includegraphics[height=11cm]{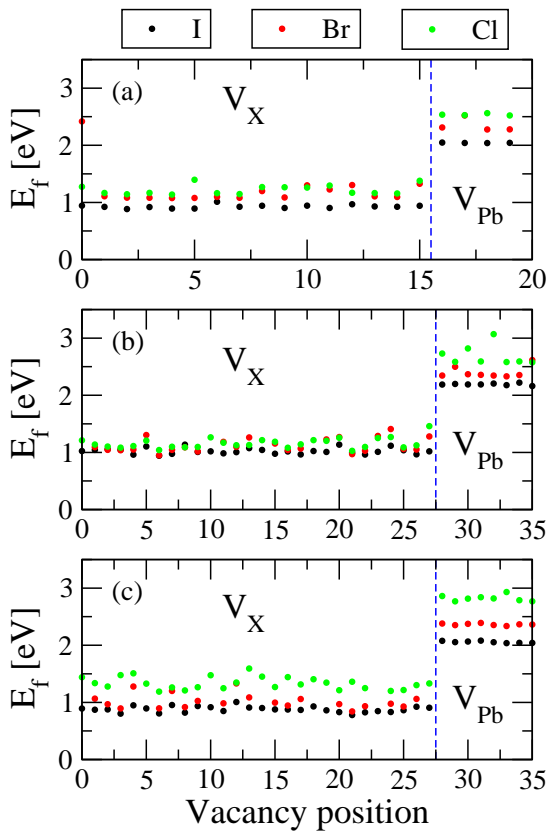}
	\caption{Formation energies of halogen  (X = I, Br, Cl) and Pb single vacancies, for the 2D perovskite classes: (a) (PA)$_2$PbX$_4$, (b) (PA)$_2$(MA)Pb$_2$X$_7$ and (c) (PA)$_2$(GA)Pb$_2$X$_7$.}
  \label{Eformation}
\end{figure}

With respect to changing the large cations in the sequence (BA,PA,HXA) the modifications of the relative stability are significantly smaller. However, replacing MA with other small cations (FA,EA,GA) leads to a larger stability enhancement. This can be explained by considering the deprotonation energies of the small cations according to the reaction: $\tilde{\text{A}}'$X $\rightarrow$ A$'$ + HX. The deprotonation energies for the small cations are 0.55 eV (MA), 1.41 eV (FA), 1.24 eV (EA) and 2.09 eV (GA), while for all three large cations (BA,PA,HXA) the value is almost the same, $\sim$0.75 eV. The large value obtained for GA, followed by FA and EA is correlated to the higher stability of the respective 2D perovskite compounds, for both intrinsic and extrinsic mechanisms. Given the fact that the deprotonation energies of the large cations are similar, the emerging picture shows the primary role of small cations in the degradation process. However, our analysis does not account for water-induced degradation, where the hydrophobic moieties of the large cations would have a significant impact.

The possibility of inducing point defects is yet another factor to be considered in the broader context of perovskite degradation, as ion migration is facilitated by vacancies or more complex defects. We investigate the formation energies of single-vacancies of iodine and lead, observing the trends induced by the halogens and small cations. The formation energy, as defined in Eq.\ (\ref{Eform}), is correlated with the likelihood of producing a certain type of defect. Figure\ \ref{Eformation} shows the formation energies for (PA)$_2$PbX$_4$, (PA)$_2$(MA)Pb$_2$X$_7$ and (PA)$_2$(GA)Pb$_2$X$_7$ classes of compounds, where all halogen and lead vacancy sites have been considered. Although there is a small spreading for each halogen type, the average formation energies indicate a clear trend: $E_{\rm f}^{\rm Cl} > E_{\rm f}^{\rm Br} > E_{\rm f}^{\rm I}$. This trend becomes more clear for GA containing perovskites, with higher formation energies, as compared to MA or the X$_4$ class. Moreover, the same trend is observed for Pb vacancies, i.e. bromine and chlorine based perovskites would enhance the Pb vacancy formation energy as well.  


\section{Experimental validation}

The experimental validation is performed by comparing electronic and stability properties of compounds from (PA)$_2$PbX$_4$ and (PA)$_2$(MA)Pb$_2$X$_7$ classes (X = I, Br, Cl), with the results obtained from {\it ab initio} calculations.

\subsection{Sample characterization}

\begin{figure}[t]
\centering
  \includegraphics[height=6.0cm]{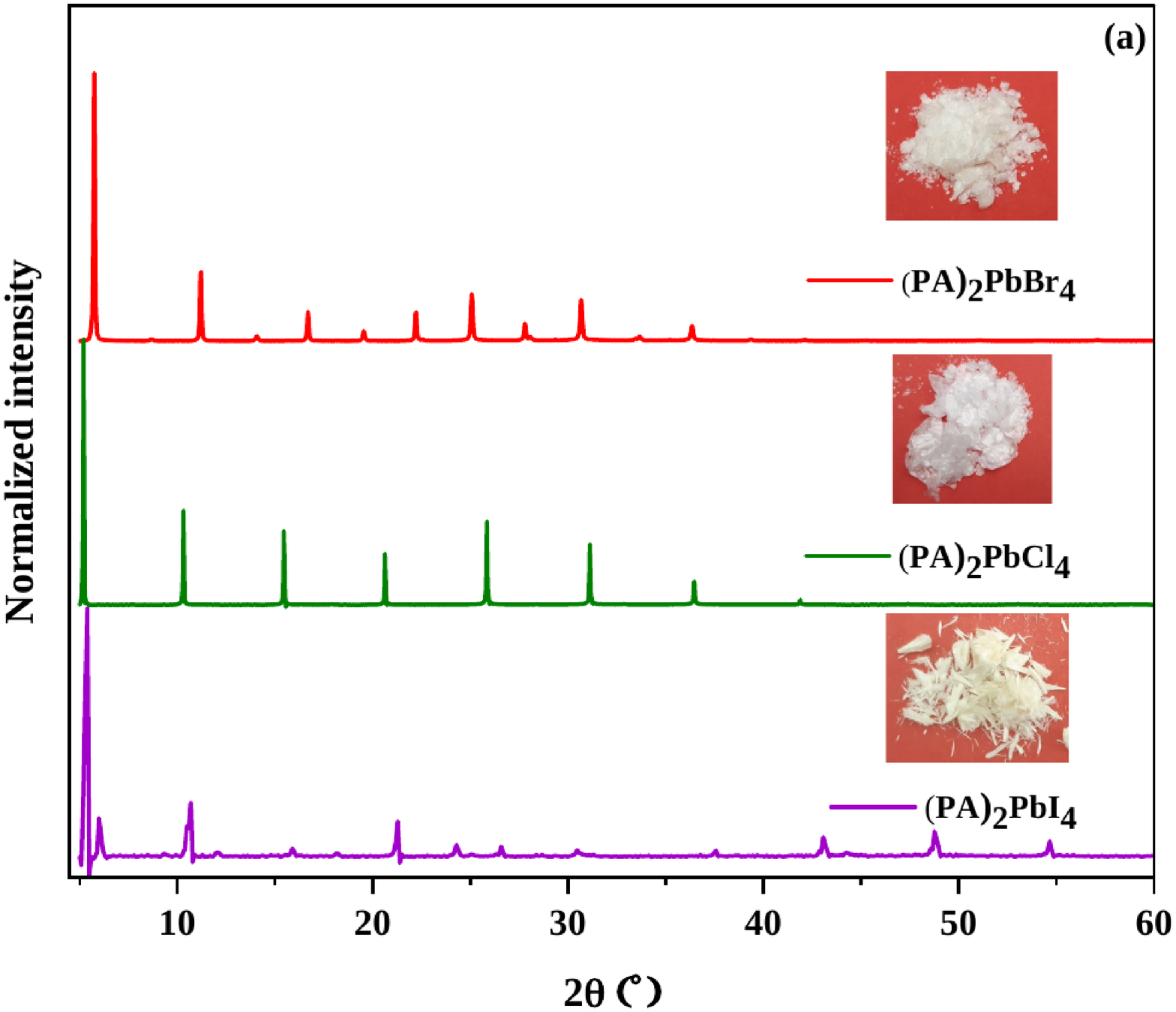}\\
  \includegraphics[height=6.0cm]{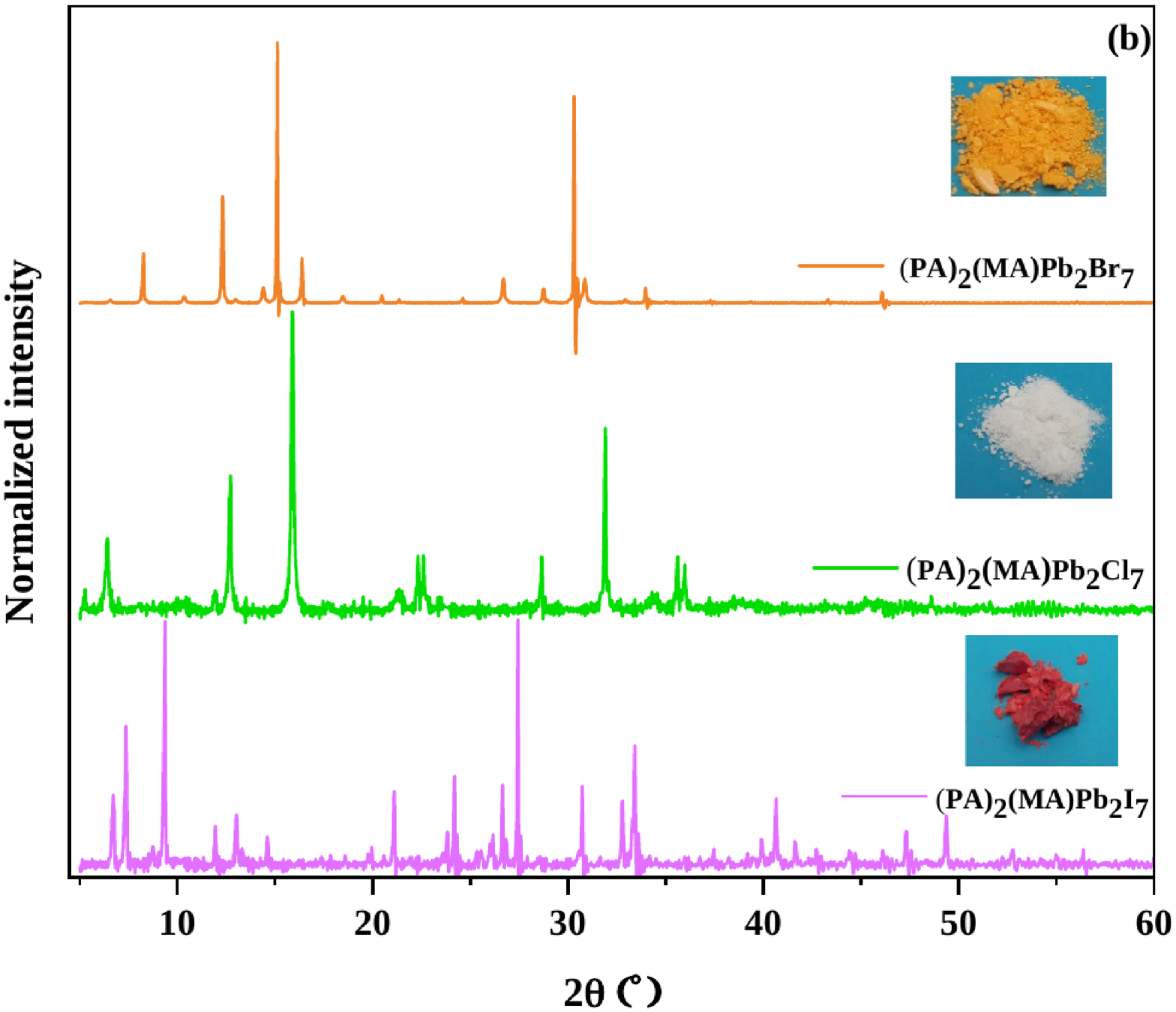}
	\caption{XRD patterns of synthesized quasi-2D perovskites: (a) (PA)$_2$PbX$_4$ and (b) (PA)$_2$(MA)Pb$_2$X$_7$, X = I, Br, Cl. The typical 2D perovskite lines are present for both classes of perovskite compounds.}
  \label{XRD}
\end{figure}

To confirm the structure of the synthesized (PA)$_2$PbX$_4$ and (PA)$_2$(MA)Pb$_2$X$_7$ compounds, X-ray diffraction has been carried out and shown in Fig.\ \ref{XRD}. The diffractograms of the (PA)$_2$PbX$_4$ class, gathered in Fig.\ \ref{XRD}(a), confirm the formation of 2D perovskite crystals. The crystals present specific diffraction lines for 2D perovskites at 5.74$^\circ$, 11.12$^\circ$, 16.68$^\circ$, 22.17$^\circ$, 27.78$^\circ$, 30.60$^\circ$ and 36.35$^\circ$ in accordance to the diffraction peaks reported in the literature \cite{C8TC06129G,doi:10.1063/1.5023797}. For the (PA)$_2$(MA)Pb$_2$X$_7$ compounds, two phases are presented in the synthesized crystals as depicted in Fig.\ \ref{XRD}(b). The 2D RP perovskite with $n=2$ phase is dominant, but also diffraction peaks of 2D layered perovskite with $n=1$ phase are observed. This is consistent with other reported studies in the literature \cite{min2020fabrication,chang2018facile}.

\begin{figure*}[t]
 \centering
 \includegraphics[height=4.5cm]{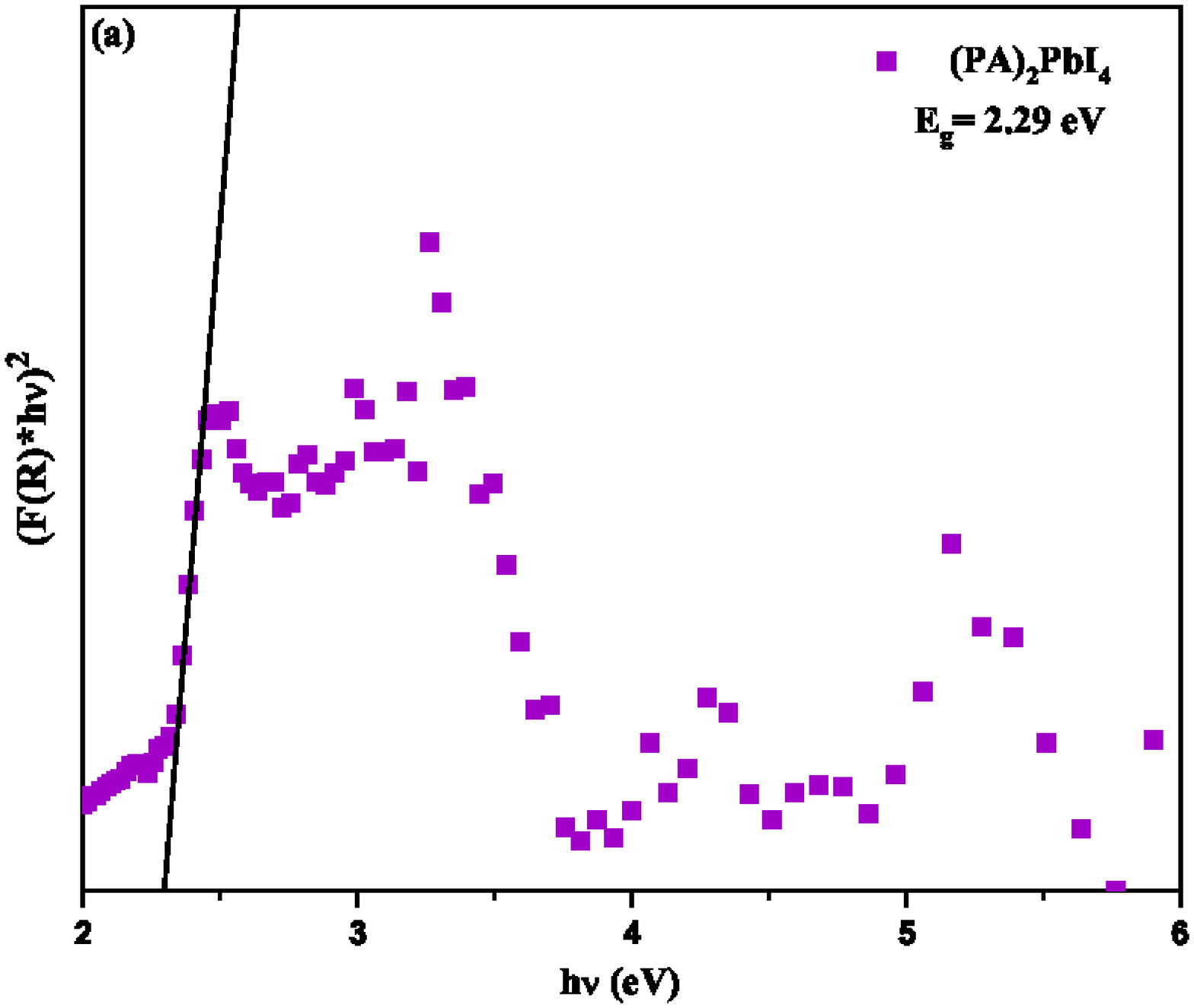}\hspace{0.25cm}
 \includegraphics[height=4.5cm]{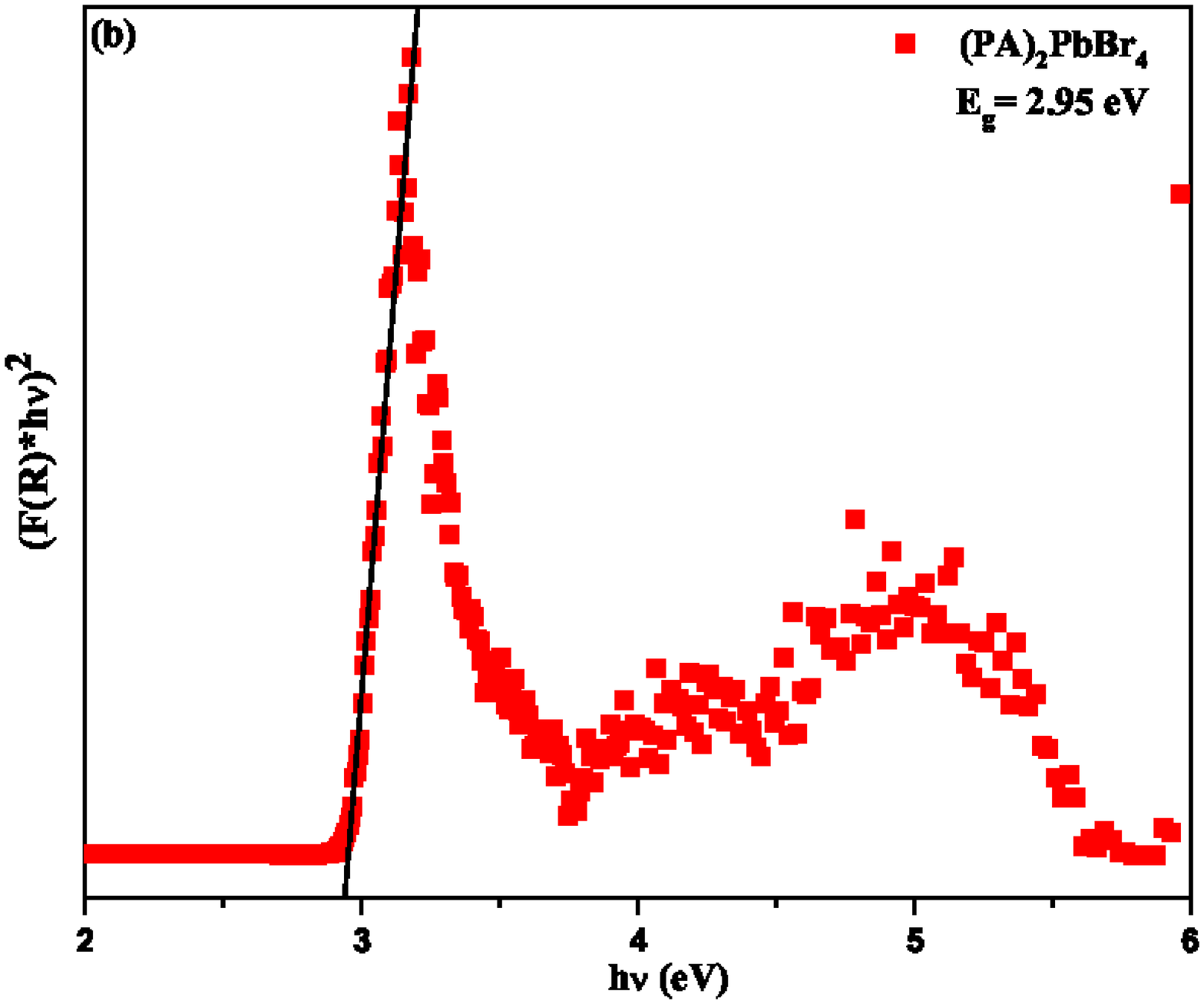}\hspace{0.25cm}
 \includegraphics[height=4.5cm]{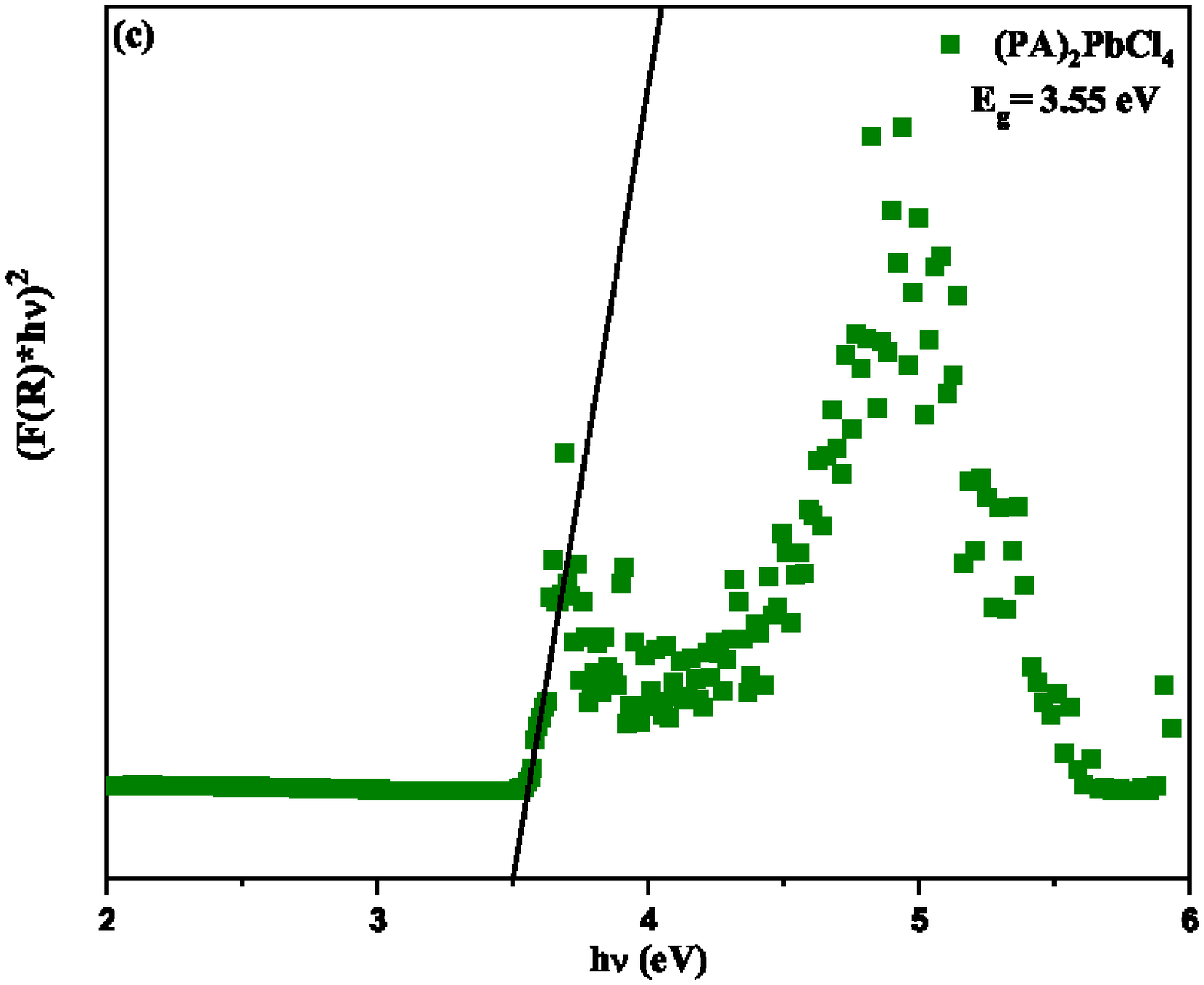} \vspace*{0.1cm}\\
 \includegraphics[height=4.7cm]{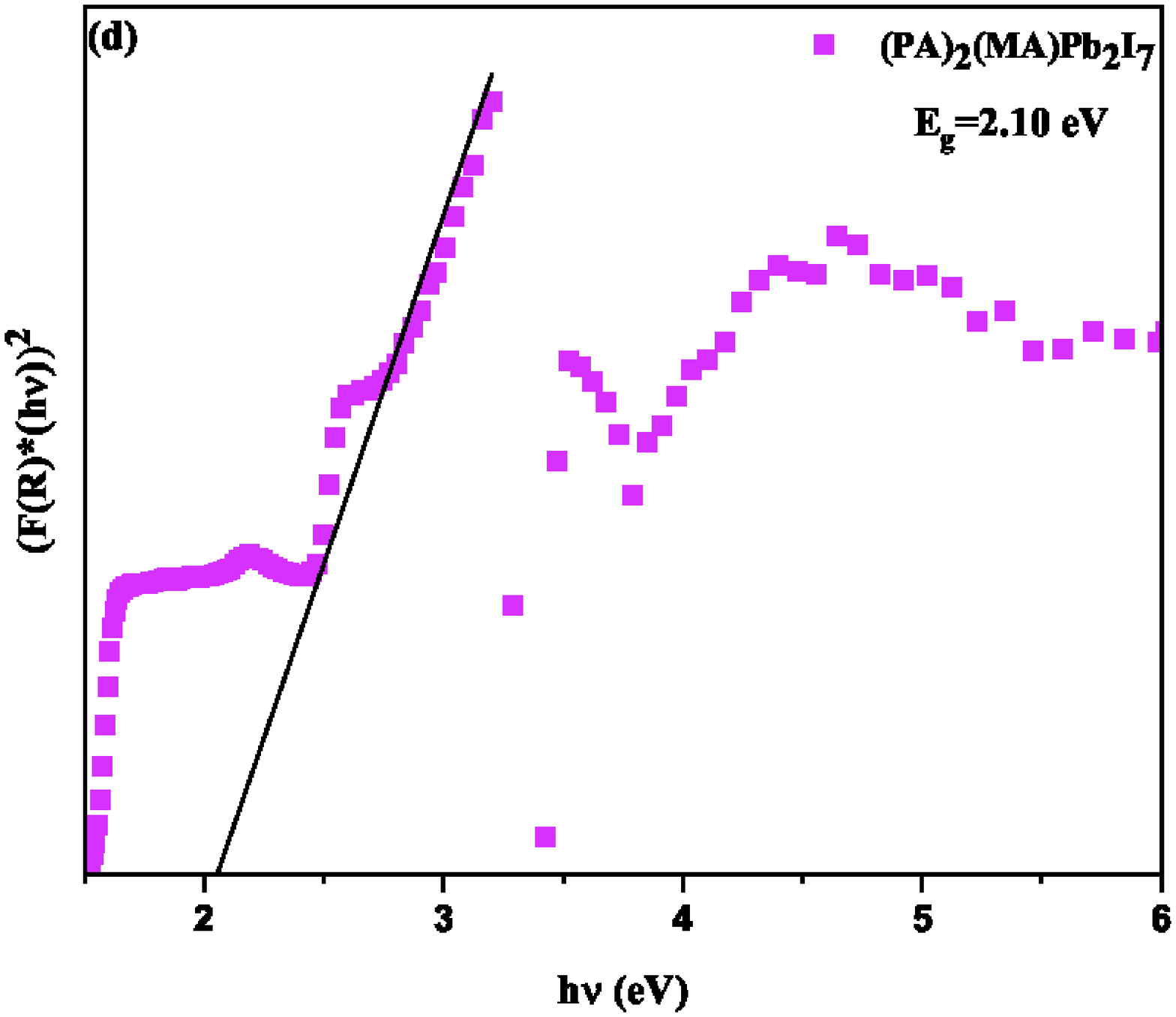}\hspace{0.25cm}
 \includegraphics[height=4.7cm]{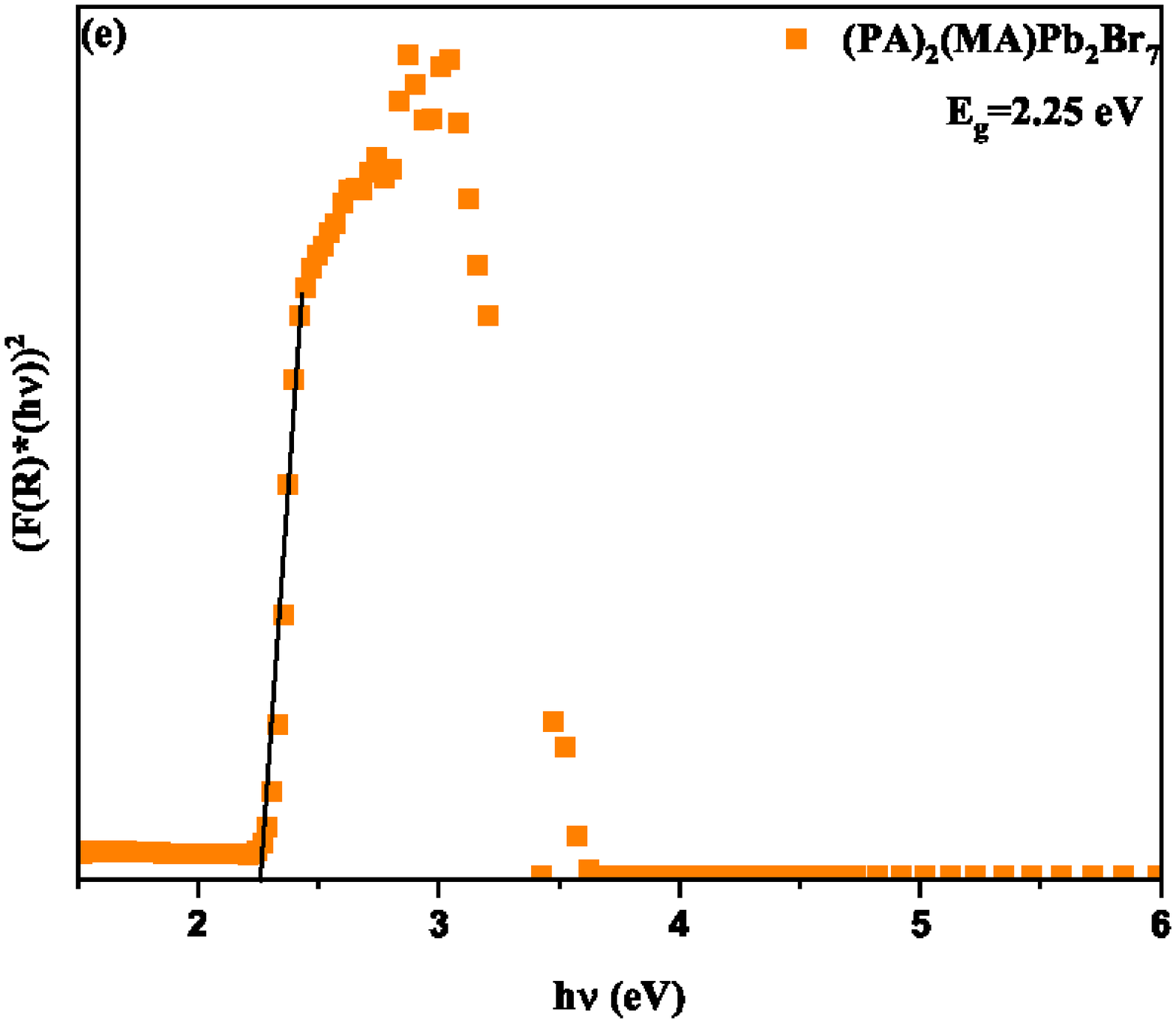}\hspace{0.25cm}
 \includegraphics[height=4.7cm]{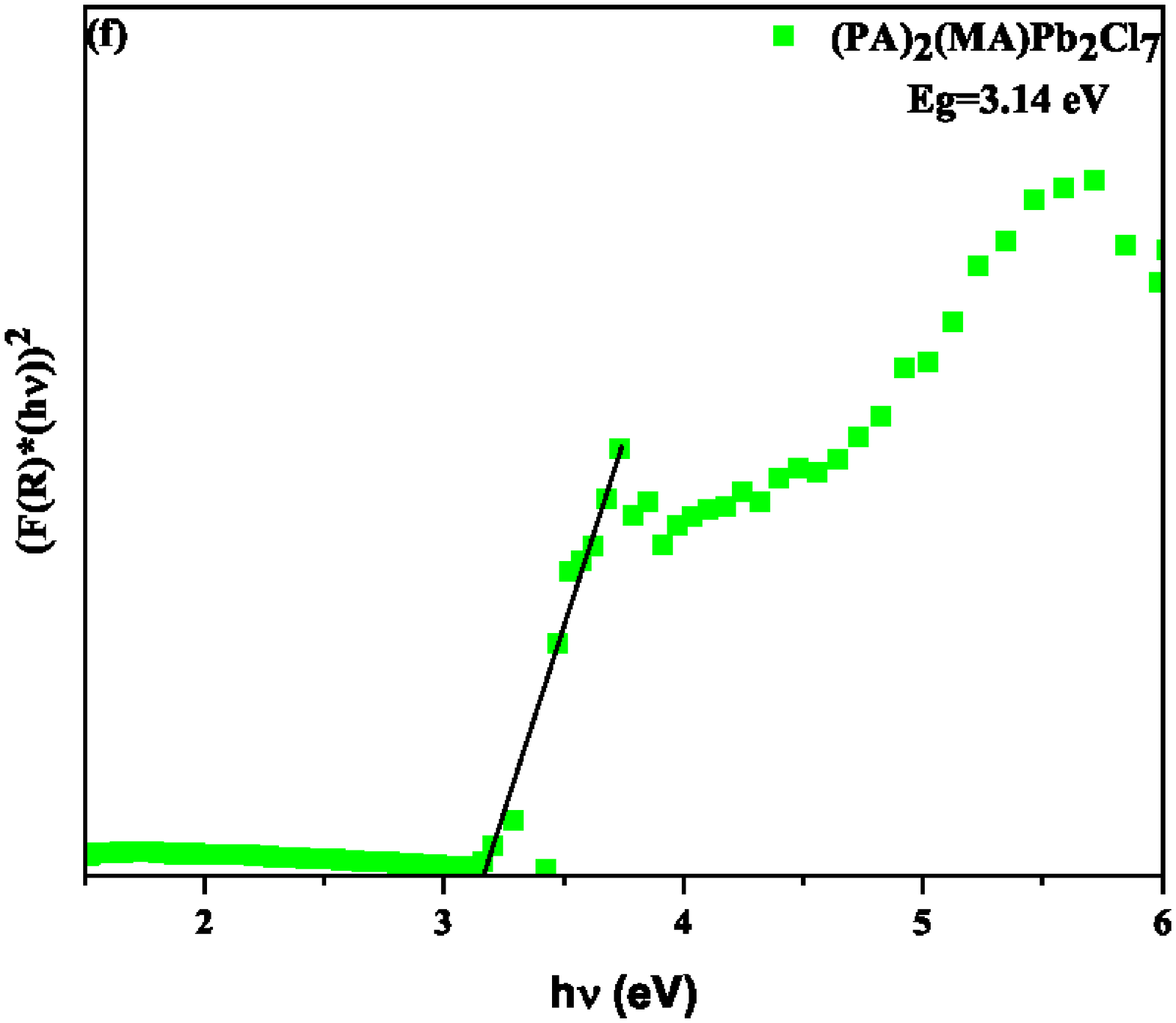}\\
	\caption{Measured optical absorption for (PA)$_2$PbX$_4$ (a,b,c) and (PA)$_2$(MA)Pb$_2$I$_7$ (d,e,f),  with X = I, Br, Cl. The band gap energies are in close correlation with the values obtained by numerical simulations, shown in Figs.\ \ref{A-MA-X7} and \ref{DOS_PA-X4-X10}. }
 \label{OptAbs}
\end{figure*}

The estimated band gaps for the synthesized crystals are indicated in Figs.\ \ref{OptAbs}(a,b) and are in good agreement with the obtained theoretical values: (PA)$_2$PbX$_4$ (DFT values in brackets) -- 2.29 (1.92) eV, 2.95 (2.46) eV, 3.55 (3.14) eV; (PA)$_2$MAPb$_2$X$_7$ -- 2.1 (1.74) eV, 2.25 (2.29) eV, 3.14 (2.62) eV.

\begin{figure}[h]
\centering
  \includegraphics[height=7.0cm]{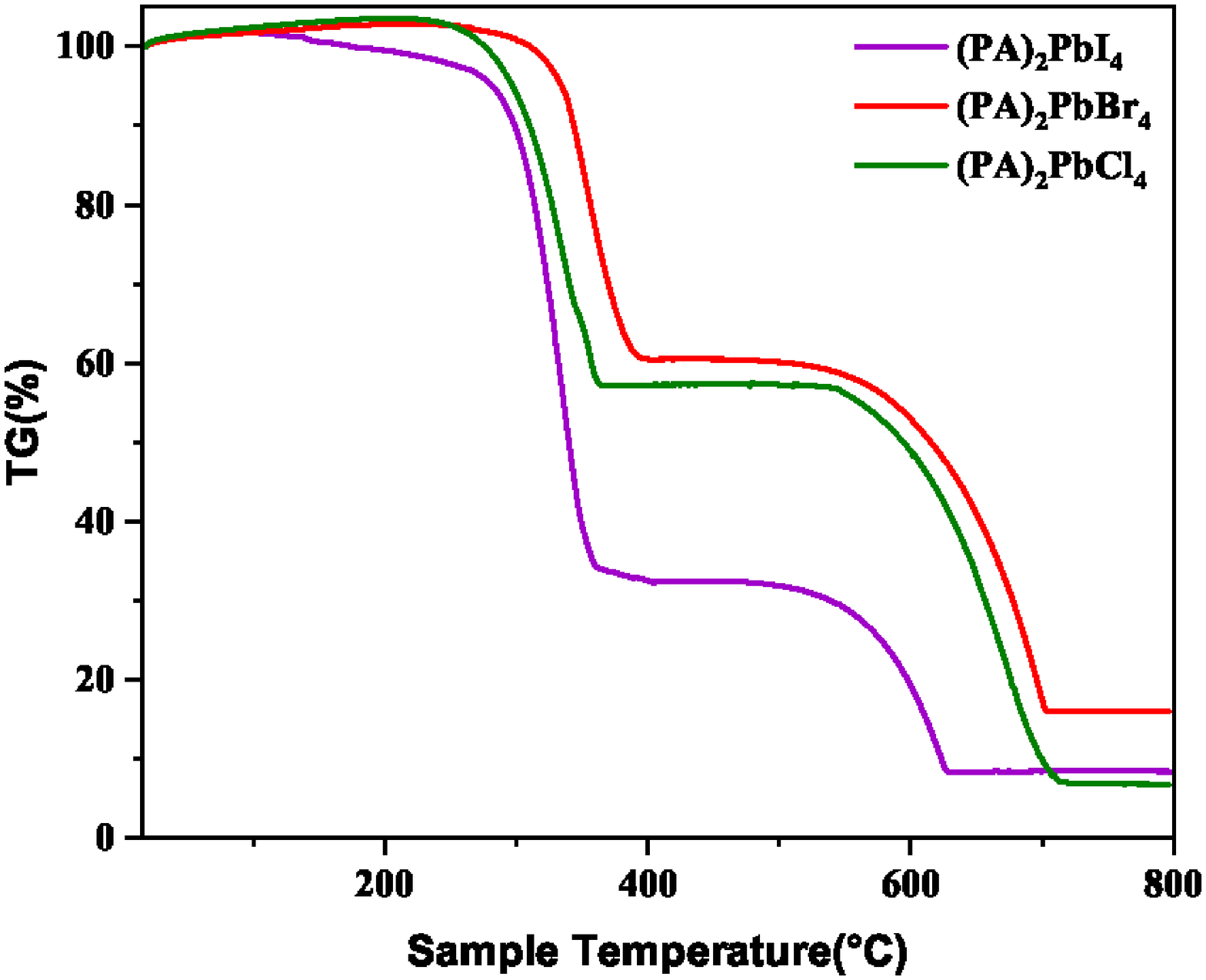}
	\caption{Thermogravimetric measurements performed for (PA)$_2$PbX$_4$ (X = I, Br, Cl) compounds in inert (N$_2$) atmosphere, revealing the increase of the stability in the halogen sequence as I $\nearrow$ Cl $\nearrow$ Br. }
  \label{TG}
\end{figure}

The intrinsic degradation mechanism assumes that the decomposition of the perovskite is triggered without any foreign species being involved. This can occur at elevated temperatures or bias stress, conditions that can be found during the operation of the solar cells. In order to simulate a degradation effect, we perform thermogravimetric measurements, allowing the perovskite compounds to fully decompose into constituents.
Thermal analysis of the (PA)$_2$PbX$_4$ crystals was performed under nitrogen atmosphere in the 20--800 $^\circ$C range and the obtained results are presented in Fig.\ \ref{TG} . The thermogravimetric profile of the (PA)$_2$PbI$_4$ sample shows the largest weight loss of $\sim$77\%, which can be noticed from a temperature as low as 100 $^\circ$C and up to 368 $^\circ$C accompanied by an endothermic process, probably due to the decomposition of organic cation. Then the remnant material is stable up to 533 $^\circ$C. At this temperature, a weight loss of $\sim$85\% can be observed until the temperature reaches 627 $^\circ$C. Similar decomposition processes are found for the other two perovskite compounds, (PA)$_2$PbBr$_4$ and (PA)$_2$PbCl$_4$, albeit the TG curves are shifted towards higher temperature values.

For all three  (PA)$_2$PbX$_4$ perovskites the process of decomposition starts at different on-set temperatures, where the mass begins to decrease below the initial value: 180 $^\circ$C, 270 $^\circ$C and 310 $^\circ$C for X = I, Cl, Br, respectively. The decomposition curves have a similar shape, in two steps, the first one showing the decomposition of the organic cation, followed by the decomposition of PbX$_2$. Their relative shift implies that the (PA)$_2$PbBr$_4$ is the most stable amongst the $X_4$ group, followed by (PA)$_2$PbCl$_4$ and (PA)$_2$PbI$_4$. This is in accordance with the theoretical prediction regarding the intrinsic degradation mechanism, as described by Eq. (\ref{int_degr}).  
The stability of (PA)$_2$PbBr$_4$ and (PA)$_2$PbCl$_4$ was also studied in O$_2$ atmosphere. The decomposition process starts with 20 $^\circ$C earlier for the compound based on Br and with 10 $^\circ$C for the one based on Cl. These data
indicate that the compound containing Cl gains stability with respect to extrinsic degradation, in agreement with theoretical data.

Given the electronic and degradation trends, one may conclude that halogens have a strong influence on both the absorption efficiency and stability. On one hand, increasing the atomic number of the halogen, the band gap decreases, making it more suitable for absorbing within the solar spectrum, while, on the other hand, the stability is reduced. Therefore, mixed halide perovskites of type I$_x$Br$_{1-x}$ or I$_x$Cl$_{1-x}$ can be a good compromise for enhanced solar cell efficiency and stability, compared to monohalogen compounds. Even though the organic cation has a smaller influence on the band gap, DFT calculations suggest that the stability can be further improved by using proper small cation mixtures.


\subsection{Experimental methods}

The crystals of quasi-2D perovskites were obtained by anti-solvent vapor-assisted crystallization (AVC), where the perovskite solutions were prepared using a DMF/DMSO solvent mixture and dichloromethane (DCM) has been used as anti-solvent for the crystallization. In the AVC procedure, the corresponding alkylammonium salt and the lead salts were dissolved stoichiometrically in the appropriate volume of solvent. The solution was placed in a closed vial containing DCM as anti-solvent in order to crystallize the species in solution. The structure of the synthesized crystals was investigated by X-ray diffraction (XRD), optical absorption and thermogravimetric (TG) measurements.

The XRD spectra were recorded with a Bruker-AXS D8 ADVANCE diffractometer using a Cu-$K_{\alpha}$ (0.1541 nm) radiation and LynxEye detector. The diffractograms were recorded over a 2$\theta$ range of 5--80 degrees at a scanning speed of 0.03 degrees$\cdot$min$^{-1}$ and a counting time of 0.5 s per point.

The optical properties of the crystals were investigated using DR-UV-Vis spectroscopy (AvaLight-Xe pulsed light source connected to an AvaSpec-ULS 2048L spectrometer equipped with an AvaSphere 80-REFL integration sphere). The spectra were recorded in the range of 200-1000 nm using BaSO$_4$ as baseline.

The optical band gap ($E_g$) of the prepared crystals was calculated using the Kubelka-Munk equation in terms of reflectance ($R$):
\begin{equation}
F(R) = \frac{(1-R)^2}{2R}
\label{FR}
\end{equation}
where $R$ is the diffuse reflectance of the examined sample and $F(R)$ is called the remission or Kubelka-Munk function. Thus, the band gap was obtained using the following equation:
\begin{equation}
(F(R) \cdot h\nu)^2 = A(h\nu - E_g)
\label{FR_Eg}
\end{equation}

The thermogravimetric analysis of the perovskite compounds was carried out with a Themys One 1150 TGA in nitrogen atmosphere, at a heating rate of 10 K/min.

\section{Conclusions}

The optoelectronic properties and stability of quasi-2D perovskites with large alkylammonium cations have been evaluated by {\it ab initio} calculations and validated by experiment. The electronic structure analysis shows an increase of the band gap with the size of the large cation and, more substantially, with decreasing the halogen atomic number. Although 2D perovskites already possess larger than optimal band gaps due to the quantum confinement, it is found that the compounds based on bromine and chlorine have larger stability with respect to the intrinsic and extrinsic mechanisms. The best compromise between these two opposing trends for the efficiency and stability can be achieved if further research reveals the optimal mixture of halogens and cations. Changing the small cations (MA,FA,EA,GA) has shown little gap variations, while the stability can be enhanced, as in the the case of MA-GA 2D perovskites. The main trends concerning the optoelectronic properties are validated experimentally for (PA)$_2$PbX$_4$ and (PA)$_2$(MA)Pb$_2$X$_7$ classes of compounds, albeit the measured energy gaps are systematically larger, as expected, due to the typical DFT underestimation. Furthermore, thermogravimetric measurements show that (PA)$_2$PbBr$_4$ is most stable, followed by chlorine and iodine based counterpart. This outlines the halogen influence on the stability, being in agreement with the theoretical predictions obtained for the intrinsic degradation mechanism.

\section*{Author Contributions}
N.F. and G.A.N. performed the {\it ab initio} calculations and contributed to the writing of the theoretical part. C.-A. P.-S. contributed to writing, review and editing. D.V.A and A.M. contributed to review and editing.  A.G.M., S.D. performed the experiments and the corresponding analyses and contributed to the writing, review and editing. M.F. and I.P. formulated the concepts and methodology for the the experimental work and contributed to the writing of the experimental part.

\section*{Conflicts of interest}
There are no conflicts to declare.

\section*{Acknowledgements}
This work was supported by a grant of the Romanian Ministry of Research, Innovation and Digitalization, CCCDI - UEFISCDI, project number PN-III-P2-2.1-PED-2019-1567, within PNCDI III. All NIMP authors acknowledge partial financial support from Core Program PN19-03 (contract no. 21N/08.02.2019).

\bibliography{manuscript} 

\begin{thebibliography}{32}%
\makeatletter
\providecommand \@ifxundefined [1]{%
 \@ifx{#1\undefined}
}%
\providecommand \@ifnum [1]{%
 \ifnum #1\expandafter \@firstoftwo
 \else \expandafter \@secondoftwo
 \fi
}%
\providecommand \@ifx [1]{%
 \ifx #1\expandafter \@firstoftwo
 \else \expandafter \@secondoftwo
 \fi
}%
\providecommand \natexlab [1]{#1}%
\providecommand \enquote  [1]{``#1''}%
\providecommand \bibnamefont  [1]{#1}%
\providecommand \bibfnamefont [1]{#1}%
\providecommand \citenamefont [1]{#1}%
\providecommand \href@noop [0]{\@secondoftwo}%
\providecommand \href [0]{\begingroup \@sanitize@url \@href}%
\providecommand \@href[1]{\@@startlink{#1}\@@href}%
\providecommand \@@href[1]{\endgroup#1\@@endlink}%
\providecommand \@sanitize@url [0]{\catcode `\\12\catcode `\$12\catcode
  `\&12\catcode `\#12\catcode `\^12\catcode `\_12\catcode `\%12\relax}%
\providecommand \@@startlink[1]{}%
\providecommand \@@endlink[0]{}%
\providecommand \url  [0]{\begingroup\@sanitize@url \@url }%
\providecommand \@url [1]{\endgroup\@href {#1}{\urlprefix }}%
\providecommand \urlprefix  [0]{URL }%
\providecommand \Eprint [0]{\href }%
\providecommand \doibase [0]{http://dx.doi.org/}%
\providecommand \selectlanguage [0]{\@gobble}%
\providecommand \bibinfo  [0]{\@secondoftwo}%
\providecommand \bibfield  [0]{\@secondoftwo}%
\providecommand \translation [1]{[#1]}%
\providecommand \BibitemOpen [0]{}%
\providecommand \bibitemStop [0]{}%
\providecommand \bibitemNoStop [0]{.\EOS\space}%
\providecommand \EOS [0]{\spacefactor3000\relax}%
\providecommand \BibitemShut  [1]{\csname bibitem#1\endcsname}%
\let\auto@bib@innerbib\@empty
\bibitem [{nre()}]{nrel}%
  \BibitemOpen
  \href@noop {} {\emph {\bibinfo {title} {Best Research Cells Efficiencies from
  NREL (2022), https://www.nrel.gov/pv/cell-efficiency.html}}}\BibitemShut
  {NoStop}%
\bibitem [{\citenamefont {Chen}\ \emph {et~al.}(2015)\citenamefont {Chen},
  \citenamefont {{De Marco}}, \citenamefont {Yang}, \citenamefont {Song},
  \citenamefont {Chen}, \citenamefont {Zhao}, \citenamefont {Hong},
  \citenamefont {Zhou},\ and\ \citenamefont {Yang}}]{CHEN2015355}%
  \BibitemOpen
  \bibfield  {author} {\bibinfo {author} {\bibfnamefont {Q.}~\bibnamefont
  {Chen}}, \bibinfo {author} {\bibfnamefont {N.}~\bibnamefont {{De Marco}}},
  \bibinfo {author} {\bibfnamefont {Y.~M.}\ \bibnamefont {Yang}}, \bibinfo
  {author} {\bibfnamefont {T.-B.}\ \bibnamefont {Song}}, \bibinfo {author}
  {\bibfnamefont {C.-C.}\ \bibnamefont {Chen}}, \bibinfo {author}
  {\bibfnamefont {H.}~\bibnamefont {Zhao}}, \bibinfo {author} {\bibfnamefont
  {Z.}~\bibnamefont {Hong}}, \bibinfo {author} {\bibfnamefont {H.}~\bibnamefont
  {Zhou}}, \ and\ \bibinfo {author} {\bibfnamefont {Y.}~\bibnamefont {Yang}},\
  }\href@noop {} {\bibfield  {journal} {\bibinfo  {journal} {Nano Today}\
  }\textbf {\bibinfo {volume} {10}},\ \bibinfo {pages} {355} (\bibinfo {year}
  {2015})}\BibitemShut {NoStop}%
\bibitem [{\citenamefont {Park}\ \emph {et~al.}(2021)\citenamefont {Park},
  \citenamefont {Ali}, \citenamefont {Mall}, \citenamefont {Bensmail},
  \citenamefont {Sanvito},\ and\ \citenamefont {El-Mellouhi}}]{Park_2021}%
  \BibitemOpen
  \bibfield  {author} {\bibinfo {author} {\bibfnamefont {H.}~\bibnamefont
  {Park}}, \bibinfo {author} {\bibfnamefont {A.}~\bibnamefont {Ali}}, \bibinfo
  {author} {\bibfnamefont {R.}~\bibnamefont {Mall}}, \bibinfo {author}
  {\bibfnamefont {H.}~\bibnamefont {Bensmail}}, \bibinfo {author}
  {\bibfnamefont {S.}~\bibnamefont {Sanvito}}, \ and\ \bibinfo {author}
  {\bibfnamefont {F.}~\bibnamefont {El-Mellouhi}},\ }\href@noop {} {\bibfield
  {journal} {\bibinfo  {journal} {Machine Learning: Science and Technology}\
  }\textbf {\bibinfo {volume} {2}},\ \bibinfo {pages} {025030} (\bibinfo {year}
  {2021})}\BibitemShut {NoStop}%
\bibitem [{\citenamefont {Filipoiu}\ \emph {et~al.}(2021)\citenamefont
  {Filipoiu}, \citenamefont {Mitran}, \citenamefont {Anghel}, \citenamefont
  {Florea}, \citenamefont {Pintilie}, \citenamefont {Manolescu},\ and\
  \citenamefont {Nemnes}}]{en14175431}%
  \BibitemOpen
  \bibfield  {author} {\bibinfo {author} {\bibfnamefont {N.}~\bibnamefont
  {Filipoiu}}, \bibinfo {author} {\bibfnamefont {T.~L.}\ \bibnamefont
  {Mitran}}, \bibinfo {author} {\bibfnamefont {D.~V.}\ \bibnamefont {Anghel}},
  \bibinfo {author} {\bibfnamefont {M.}~\bibnamefont {Florea}}, \bibinfo
  {author} {\bibfnamefont {I.}~\bibnamefont {Pintilie}}, \bibinfo {author}
  {\bibfnamefont {A.}~\bibnamefont {Manolescu}}, \ and\ \bibinfo {author}
  {\bibfnamefont {G.~A.}\ \bibnamefont {Nemnes}},\ }\href@noop {} {\bibfield
  {journal} {\bibinfo  {journal} {Energies}\ }\textbf {\bibinfo {volume}
  {14}},\ \bibinfo {pages} {5431} (\bibinfo {year} {2021})}\BibitemShut
  {NoStop}%
\bibitem [{\citenamefont {Ju}\ \emph {et~al.}(2017)\citenamefont {Ju},
  \citenamefont {Dai}, \citenamefont {Ma},\ and\ \citenamefont
  {Zeng}}]{doi:10.1021/jacs.7b04219}%
  \BibitemOpen
  \bibfield  {author} {\bibinfo {author} {\bibfnamefont {M.-G.}\ \bibnamefont
  {Ju}}, \bibinfo {author} {\bibfnamefont {J.}~\bibnamefont {Dai}}, \bibinfo
  {author} {\bibfnamefont {L.}~\bibnamefont {Ma}}, \ and\ \bibinfo {author}
  {\bibfnamefont {X.~C.}\ \bibnamefont {Zeng}},\ }\href@noop {} {\bibfield
  {journal} {\bibinfo  {journal} {Journal of the American Chemical Society}\
  }\textbf {\bibinfo {volume} {139}},\ \bibinfo {pages} {8038} (\bibinfo {year}
  {2017})}\BibitemShut {NoStop}%
\bibitem [{\citenamefont {Azhari}\ \emph {et~al.}(2020)\citenamefont {Azhari},
  \citenamefont {Then}, \citenamefont {Halin}, \citenamefont {Sepeai},\ and\
  \citenamefont {Ludin}}]{Azhari_2020}%
  \BibitemOpen
  \bibfield  {author} {\bibinfo {author} {\bibfnamefont {A.~W.}\ \bibnamefont
  {Azhari}}, \bibinfo {author} {\bibfnamefont {F.~S.~X.}\ \bibnamefont {Then}},
  \bibinfo {author} {\bibfnamefont {D.~S.~C.}\ \bibnamefont {Halin}}, \bibinfo
  {author} {\bibfnamefont {S.}~\bibnamefont {Sepeai}}, \ and\ \bibinfo {author}
  {\bibfnamefont {N.~A.}\ \bibnamefont {Ludin}},\ }\href@noop {} {\bibfield
  {journal} {\bibinfo  {journal} {{IOP} Conference Series: Materials Science
  and Engineering}\ }\textbf {\bibinfo {volume} {957}},\ \bibinfo {pages}
  {012057} (\bibinfo {year} {2020})}\BibitemShut {NoStop}%
\bibitem [{\citenamefont {Mao}\ \emph {et~al.}(2019)\citenamefont {Mao},
  \citenamefont {Stoumpos},\ and\ \citenamefont
  {Kanatzidis}}]{doi:10.1021/jacs.8b10851}%
  \BibitemOpen
  \bibfield  {author} {\bibinfo {author} {\bibfnamefont {L.}~\bibnamefont
  {Mao}}, \bibinfo {author} {\bibfnamefont {C.~C.}\ \bibnamefont {Stoumpos}}, \
  and\ \bibinfo {author} {\bibfnamefont {M.~G.}\ \bibnamefont {Kanatzidis}},\
  }\href@noop {} {\bibfield  {journal} {\bibinfo  {journal} {Journal of the
  American Chemical Society}\ }\textbf {\bibinfo {volume} {141}},\ \bibinfo
  {pages} {1171} (\bibinfo {year} {2019})}\BibitemShut {NoStop}%
\bibitem [{\citenamefont {Zhang}\ \emph {et~al.}(2020)\citenamefont {Zhang},
  \citenamefont {Lu}, \citenamefont {Tong}, \citenamefont {Berry},
  \citenamefont {Beard},\ and\ \citenamefont {Zhu}}]{C9EE03757H}%
  \BibitemOpen
  \bibfield  {author} {\bibinfo {author} {\bibfnamefont {F.}~\bibnamefont
  {Zhang}}, \bibinfo {author} {\bibfnamefont {H.}~\bibnamefont {Lu}}, \bibinfo
  {author} {\bibfnamefont {J.}~\bibnamefont {Tong}}, \bibinfo {author}
  {\bibfnamefont {J.~J.}\ \bibnamefont {Berry}}, \bibinfo {author}
  {\bibfnamefont {M.~C.}\ \bibnamefont {Beard}}, \ and\ \bibinfo {author}
  {\bibfnamefont {K.}~\bibnamefont {Zhu}},\ }\href@noop {} {\bibfield
  {journal} {\bibinfo  {journal} {Energy Environ. Sci.}\ }\textbf {\bibinfo
  {volume} {13}},\ \bibinfo {pages} {1154} (\bibinfo {year}
  {2020})}\BibitemShut {NoStop}%
\bibitem [{\citenamefont {Kim}\ \emph {et~al.}(2021)\citenamefont {Kim},
  \citenamefont {Akhtar}, \citenamefont {Shin}, \citenamefont {Ameen},\ and\
  \citenamefont {Nazeeruddin}}]{KIM2021100405}%
  \BibitemOpen
  \bibfield  {author} {\bibinfo {author} {\bibfnamefont {E.-B.}\ \bibnamefont
  {Kim}}, \bibinfo {author} {\bibfnamefont {M.~S.}\ \bibnamefont {Akhtar}},
  \bibinfo {author} {\bibfnamefont {H.-S.}\ \bibnamefont {Shin}}, \bibinfo
  {author} {\bibfnamefont {S.}~\bibnamefont {Ameen}}, \ and\ \bibinfo {author}
  {\bibfnamefont {M.~K.}\ \bibnamefont {Nazeeruddin}},\ }\href@noop {}
  {\bibfield  {journal} {\bibinfo  {journal} {Journal of Photochemistry and
  Photobiology C: Photochemistry Reviews}\ }\textbf {\bibinfo {volume} {48}},\
  \bibinfo {pages} {100405} (\bibinfo {year} {2021})}\BibitemShut {NoStop}%
\bibitem [{\citenamefont {Elahi}\ \emph {et~al.}(2022)\citenamefont {Elahi},
  \citenamefont {Dastgeer}, \citenamefont {Siddiqui}, \citenamefont {Patil},
  \citenamefont {Iqbal},\ and\ \citenamefont {Sharma}}]{D1DT02991F}%
  \BibitemOpen
  \bibfield  {author} {\bibinfo {author} {\bibfnamefont {E.}~\bibnamefont
  {Elahi}}, \bibinfo {author} {\bibfnamefont {G.}~\bibnamefont {Dastgeer}},
  \bibinfo {author} {\bibfnamefont {A.~S.}\ \bibnamefont {Siddiqui}}, \bibinfo
  {author} {\bibfnamefont {S.~A.}\ \bibnamefont {Patil}}, \bibinfo {author}
  {\bibfnamefont {M.~W.}\ \bibnamefont {Iqbal}}, \ and\ \bibinfo {author}
  {\bibfnamefont {P.~R.}\ \bibnamefont {Sharma}},\ }\href@noop {} {\bibfield
  {journal} {\bibinfo  {journal} {Dalton Trans.}\ }\textbf {\bibinfo {volume}
  {51}},\ \bibinfo {pages} {797} (\bibinfo {year} {2022})}\BibitemShut
  {NoStop}%
\bibitem [{\citenamefont {Smith}\ \emph {et~al.}(2014)\citenamefont {Smith},
  \citenamefont {Hoke}, \citenamefont {Solis-Ibarra}, \citenamefont {McGehee},\
  and\ \citenamefont {Karunadasa}}]{https://doi.org/10.1002/anie.201406466}%
  \BibitemOpen
  \bibfield  {author} {\bibinfo {author} {\bibfnamefont {I.~C.}\ \bibnamefont
  {Smith}}, \bibinfo {author} {\bibfnamefont {E.~T.}\ \bibnamefont {Hoke}},
  \bibinfo {author} {\bibfnamefont {D.}~\bibnamefont {Solis-Ibarra}}, \bibinfo
  {author} {\bibfnamefont {M.~D.}\ \bibnamefont {McGehee}}, \ and\ \bibinfo
  {author} {\bibfnamefont {H.~I.}\ \bibnamefont {Karunadasa}},\ }\href@noop {}
  {\bibfield  {journal} {\bibinfo  {journal} {Angewandte Chemie International
  Edition}\ }\textbf {\bibinfo {volume} {53}},\ \bibinfo {pages} {11232}
  (\bibinfo {year} {2014})}\BibitemShut {NoStop}%
\bibitem [{\citenamefont {Stoumpos}\ \emph {et~al.}(2016)\citenamefont
  {Stoumpos}, \citenamefont {Cao}, \citenamefont {Clark}, \citenamefont
  {Young}, \citenamefont {Rondinelli}, \citenamefont {Jang}, \citenamefont
  {Hupp},\ and\ \citenamefont
  {Kanatzidis}}]{doi:10.1021/acs.chemmater.6b00847}%
  \BibitemOpen
  \bibfield  {author} {\bibinfo {author} {\bibfnamefont {C.~C.}\ \bibnamefont
  {Stoumpos}}, \bibinfo {author} {\bibfnamefont {D.~H.}\ \bibnamefont {Cao}},
  \bibinfo {author} {\bibfnamefont {D.~J.}\ \bibnamefont {Clark}}, \bibinfo
  {author} {\bibfnamefont {J.}~\bibnamefont {Young}}, \bibinfo {author}
  {\bibfnamefont {J.~M.}\ \bibnamefont {Rondinelli}}, \bibinfo {author}
  {\bibfnamefont {J.~I.}\ \bibnamefont {Jang}}, \bibinfo {author}
  {\bibfnamefont {J.~T.}\ \bibnamefont {Hupp}}, \ and\ \bibinfo {author}
  {\bibfnamefont {M.~G.}\ \bibnamefont {Kanatzidis}},\ }\href@noop {}
  {\bibfield  {journal} {\bibinfo  {journal} {Chemistry of Materials}\ }\textbf
  {\bibinfo {volume} {28}},\ \bibinfo {pages} {2852} (\bibinfo {year}
  {2016})}\BibitemShut {NoStop}%
\bibitem [{\citenamefont {Ortiz-Cervantes}\ \emph {et~al.}(2019)\citenamefont
  {Ortiz-Cervantes}, \citenamefont {Carmona-Monroy},\ and\ \citenamefont
  {Solis-Ibarra}}]{https://doi.org/10.1002/cssc.201802992}%
  \BibitemOpen
  \bibfield  {author} {\bibinfo {author} {\bibfnamefont {C.}~\bibnamefont
  {Ortiz-Cervantes}}, \bibinfo {author} {\bibfnamefont {P.}~\bibnamefont
  {Carmona-Monroy}}, \ and\ \bibinfo {author} {\bibfnamefont {D.}~\bibnamefont
  {Solis-Ibarra}},\ }\href@noop {} {\bibfield  {journal} {\bibinfo  {journal}
  {ChemSusChem}\ }\textbf {\bibinfo {volume} {12}},\ \bibinfo {pages} {1560}
  (\bibinfo {year} {2019})}\BibitemShut {NoStop}%
\bibitem [{\citenamefont {Maheshwari}\ \emph {et~al.}(2018)\citenamefont
  {Maheshwari}, \citenamefont {Savenije}, \citenamefont {Renaud},\ and\
  \citenamefont {Grozema}}]{doi:10.1021/acs.jpcc.8b05715}%
  \BibitemOpen
  \bibfield  {author} {\bibinfo {author} {\bibfnamefont {S.}~\bibnamefont
  {Maheshwari}}, \bibinfo {author} {\bibfnamefont {T.~J.}\ \bibnamefont
  {Savenije}}, \bibinfo {author} {\bibfnamefont {N.}~\bibnamefont {Renaud}}, \
  and\ \bibinfo {author} {\bibfnamefont {F.~C.}\ \bibnamefont {Grozema}},\
  }\href@noop {} {\bibfield  {journal} {\bibinfo  {journal} {The Journal of
  Physical Chemistry C}\ }\textbf {\bibinfo {volume} {122}},\ \bibinfo {pages}
  {17118} (\bibinfo {year} {2018})}\BibitemShut {NoStop}%
\bibitem [{\citenamefont {Mahal}\ \emph {et~al.}(2022)\citenamefont {Mahal},
  \citenamefont {Mandal},\ and\ \citenamefont
  {Pathak}}]{doi:10.1021/acs.jpcc.2c01840}%
  \BibitemOpen
  \bibfield  {author} {\bibinfo {author} {\bibfnamefont {E.}~\bibnamefont
  {Mahal}}, \bibinfo {author} {\bibfnamefont {S.~C.}\ \bibnamefont {Mandal}}, \
  and\ \bibinfo {author} {\bibfnamefont {B.}~\bibnamefont {Pathak}},\
  }\href@noop {} {\bibfield  {journal} {\bibinfo  {journal} {The Journal of
  Physical Chemistry C}\ }\textbf {\bibinfo {volume} {126}},\ \bibinfo {pages}
  {9937} (\bibinfo {year} {2022})}\BibitemShut {NoStop}%
\bibitem [{\citenamefont {Lee}\ \emph {et~al.}(2022)\citenamefont {Lee},
  \citenamefont {Lee}, \citenamefont {Kim}, \citenamefont {Park},\ and\
  \citenamefont {Sohn}}]{Lee2022}%
  \BibitemOpen
  \bibfield  {author} {\bibinfo {author} {\bibfnamefont {B.~D.}\ \bibnamefont
  {Lee}}, \bibinfo {author} {\bibfnamefont {J.-W.}\ \bibnamefont {Lee}},
  \bibinfo {author} {\bibfnamefont {M.}~\bibnamefont {Kim}}, \bibinfo {author}
  {\bibfnamefont {W.~B.}\ \bibnamefont {Park}}, \ and\ \bibinfo {author}
  {\bibfnamefont {K.-S.}\ \bibnamefont {Sohn}},\ }\href@noop {} {\bibfield
  {journal} {\bibinfo  {journal} {npj Computational Materials}\ }\textbf
  {\bibinfo {volume} {8}},\ \bibinfo {pages} {83} (\bibinfo {year}
  {2022})}\BibitemShut {NoStop}%
\bibitem [{\citenamefont {Hong}\ \emph {et~al.}(2021)\citenamefont {Hong},
  \citenamefont {Prendergast},\ and\ \citenamefont
  {Tan}}]{doi:10.1021/acs.nanolett.0c03468}%
  \BibitemOpen
  \bibfield  {author} {\bibinfo {author} {\bibfnamefont {J.}~\bibnamefont
  {Hong}}, \bibinfo {author} {\bibfnamefont {D.}~\bibnamefont {Prendergast}}, \
  and\ \bibinfo {author} {\bibfnamefont {L.~Z.}\ \bibnamefont {Tan}},\
  }\href@noop {} {\bibfield  {journal} {\bibinfo  {journal} {Nano Letters}\
  }\textbf {\bibinfo {volume} {21}},\ \bibinfo {pages} {182} (\bibinfo {year}
  {2021})}\BibitemShut {NoStop}%
\bibitem [{\citenamefont {Zhai}\ \emph {et~al.}(2017)\citenamefont {Zhai},
  \citenamefont {Baniya}, \citenamefont {Zhang}, \citenamefont {Li},
  \citenamefont {Haney}, \citenamefont {Sheng}, \citenamefont {Ehrenfreund},\
  and\ \citenamefont {Vardeny}}]{doi:10.1126/sciadv.1700704}%
  \BibitemOpen
  \bibfield  {author} {\bibinfo {author} {\bibfnamefont {Y.}~\bibnamefont
  {Zhai}}, \bibinfo {author} {\bibfnamefont {S.}~\bibnamefont {Baniya}},
  \bibinfo {author} {\bibfnamefont {C.}~\bibnamefont {Zhang}}, \bibinfo
  {author} {\bibfnamefont {J.}~\bibnamefont {Li}}, \bibinfo {author}
  {\bibfnamefont {P.}~\bibnamefont {Haney}}, \bibinfo {author} {\bibfnamefont
  {C.-X.}\ \bibnamefont {Sheng}}, \bibinfo {author} {\bibfnamefont
  {E.}~\bibnamefont {Ehrenfreund}}, \ and\ \bibinfo {author} {\bibfnamefont
  {Z.~V.}\ \bibnamefont {Vardeny}},\ }\href@noop {} {\bibfield  {journal}
  {\bibinfo  {journal} {Science Advances}\ }\textbf {\bibinfo {volume} {3}},\
  \bibinfo {pages} {e1700704} (\bibinfo {year} {2017})}\BibitemShut {NoStop}%
\bibitem [{\citenamefont {Billing}\ and\ \citenamefont
  {Lemmerer}(2007)}]{billing2007}%
  \BibitemOpen
  \bibfield  {author} {\bibinfo {author} {\bibfnamefont {D.~G.}\ \bibnamefont
  {Billing}}\ and\ \bibinfo {author} {\bibfnamefont {A.}~\bibnamefont
  {Lemmerer}},\ }\href@noop {} {\bibfield  {journal} {\bibinfo  {journal} {Acta
  crystallographica}\ }\textbf {\bibinfo {volume} {63}},\ \bibinfo {pages}
  {735} (\bibinfo {year} {2007})}\BibitemShut {NoStop}%
\bibitem [{\citenamefont {Wygant}\ \emph {et~al.}(2020)\citenamefont {Wygant},
  \citenamefont {Ye}, \citenamefont {Dolocan},\ and\ \citenamefont
  {Mullins}}]{doi:10.1021/acs.jpcc.0c02822}%
  \BibitemOpen
  \bibfield  {author} {\bibinfo {author} {\bibfnamefont {B.~R.}\ \bibnamefont
  {Wygant}}, \bibinfo {author} {\bibfnamefont {A.~Z.}\ \bibnamefont {Ye}},
  \bibinfo {author} {\bibfnamefont {A.}~\bibnamefont {Dolocan}}, \ and\
  \bibinfo {author} {\bibfnamefont {C.~B.}\ \bibnamefont {Mullins}},\
  }\href@noop {} {\bibfield  {journal} {\bibinfo  {journal} {The Journal of
  Physical Chemistry C}\ }\textbf {\bibinfo {volume} {124}},\ \bibinfo {pages}
  {10887} (\bibinfo {year} {2020})}\BibitemShut {NoStop}%
\bibitem [{\citenamefont {Yao}\ \emph {et~al.}(2018)\citenamefont {Yao},
  \citenamefont {Zhou}, \citenamefont {Yin}, \citenamefont {Li}, \citenamefont
  {Han}, \citenamefont {Tai}, \citenamefont {Zhou}, \citenamefont {Li},
  \citenamefont {Hao},\ and\ \citenamefont {Lin}}]{C8CE00999F}%
  \BibitemOpen
  \bibfield  {author} {\bibinfo {author} {\bibfnamefont {Z.}~\bibnamefont
  {Yao}}, \bibinfo {author} {\bibfnamefont {Y.}~\bibnamefont {Zhou}}, \bibinfo
  {author} {\bibfnamefont {X.}~\bibnamefont {Yin}}, \bibinfo {author}
  {\bibfnamefont {X.}~\bibnamefont {Li}}, \bibinfo {author} {\bibfnamefont
  {J.}~\bibnamefont {Han}}, \bibinfo {author} {\bibfnamefont {M.}~\bibnamefont
  {Tai}}, \bibinfo {author} {\bibfnamefont {Y.}~\bibnamefont {Zhou}}, \bibinfo
  {author} {\bibfnamefont {J.}~\bibnamefont {Li}}, \bibinfo {author}
  {\bibfnamefont {F.}~\bibnamefont {Hao}}, \ and\ \bibinfo {author}
  {\bibfnamefont {H.}~\bibnamefont {Lin}},\ }\href@noop {} {\bibfield
  {journal} {\bibinfo  {journal} {CrystEngComm}\ }\textbf {\bibinfo {volume}
  {20}},\ \bibinfo {pages} {6704} (\bibinfo {year} {2018})}\BibitemShut
  {NoStop}%
\bibitem [{\citenamefont {Xu}\ \emph {et~al.}(2019)\citenamefont {Xu},
  \citenamefont {Li}, \citenamefont {Liu}, \citenamefont {Ji}, \citenamefont
  {Chen}, \citenamefont {Li}, \citenamefont {Han}, \citenamefont {Hong},
  \citenamefont {Luo},\ and\ \citenamefont
  {Sun}}]{https://doi.org/10.1002/adom.201900308}%
  \BibitemOpen
  \bibfield  {author} {\bibinfo {author} {\bibfnamefont {Z.}~\bibnamefont
  {Xu}}, \bibinfo {author} {\bibfnamefont {Y.}~\bibnamefont {Li}}, \bibinfo
  {author} {\bibfnamefont {X.}~\bibnamefont {Liu}}, \bibinfo {author}
  {\bibfnamefont {C.}~\bibnamefont {Ji}}, \bibinfo {author} {\bibfnamefont
  {H.}~\bibnamefont {Chen}}, \bibinfo {author} {\bibfnamefont {L.}~\bibnamefont
  {Li}}, \bibinfo {author} {\bibfnamefont {S.}~\bibnamefont {Han}}, \bibinfo
  {author} {\bibfnamefont {M.}~\bibnamefont {Hong}}, \bibinfo {author}
  {\bibfnamefont {J.}~\bibnamefont {Luo}}, \ and\ \bibinfo {author}
  {\bibfnamefont {Z.}~\bibnamefont {Sun}},\ }\href@noop {} {\bibfield
  {journal} {\bibinfo  {journal} {Advanced Optical Materials}\ }\textbf
  {\bibinfo {volume} {7}},\ \bibinfo {pages} {1900308} (\bibinfo {year}
  {2019})}\BibitemShut {NoStop}%
\bibitem [{\citenamefont {Soler}\ \emph {et~al.}(2002)\citenamefont {Soler},
  \citenamefont {Artacho}, \citenamefont {Gale}, \citenamefont {Garcia},
  \citenamefont {Junquera}, \citenamefont {Ordejon},\ and\ \citenamefont
  {Sanchez-Portal}}]{0953-8984-14-11-302}%
  \BibitemOpen
  \bibfield  {author} {\bibinfo {author} {\bibfnamefont {J.~M.}\ \bibnamefont
  {Soler}}, \bibinfo {author} {\bibfnamefont {E.}~\bibnamefont {Artacho}},
  \bibinfo {author} {\bibfnamefont {J.~D.}\ \bibnamefont {Gale}}, \bibinfo
  {author} {\bibfnamefont {A.}~\bibnamefont {Garcia}}, \bibinfo {author}
  {\bibfnamefont {J.}~\bibnamefont {Junquera}}, \bibinfo {author}
  {\bibfnamefont {P.}~\bibnamefont {Ordejon}}, \ and\ \bibinfo {author}
  {\bibfnamefont {D.}~\bibnamefont {Sanchez-Portal}},\ }\href@noop {}
  {\bibfield  {journal} {\bibinfo  {journal} {Journal of Physics: Condensed
  Matter}\ }\textbf {\bibinfo {volume} {14}},\ \bibinfo {pages} {2745}
  (\bibinfo {year} {2002})}\BibitemShut {NoStop}%
\bibitem [{\citenamefont {Ceperley}\ and\ \citenamefont
  {Alder}(1980)}]{PhysRevLett.45.566}%
  \BibitemOpen
  \bibfield  {author} {\bibinfo {author} {\bibfnamefont {D.~M.}\ \bibnamefont
  {Ceperley}}\ and\ \bibinfo {author} {\bibfnamefont {B.~J.}\ \bibnamefont
  {Alder}},\ }\href@noop {} {\bibfield  {journal} {\bibinfo  {journal} {Phys.
  Rev. Lett.}\ }\textbf {\bibinfo {volume} {45}},\ \bibinfo {pages} {566}
  (\bibinfo {year} {1980})}\BibitemShut {NoStop}%
\bibitem [{\citenamefont {Troullier}\ and\ \citenamefont
  {Martins}(1991)}]{PhysRevB.43.1993}%
  \BibitemOpen
  \bibfield  {author} {\bibinfo {author} {\bibfnamefont {N.}~\bibnamefont
  {Troullier}}\ and\ \bibinfo {author} {\bibfnamefont {J.~L.}\ \bibnamefont
  {Martins}},\ }\href@noop {} {\bibfield  {journal} {\bibinfo  {journal} {Phys.
  Rev. B}\ }\textbf {\bibinfo {volume} {43}},\ \bibinfo {pages} {1993}
  (\bibinfo {year} {1991})}\BibitemShut {NoStop}%
\bibitem [{\citenamefont {Senocrate}\ \emph {et~al.}(2019)\citenamefont
  {Senocrate}, \citenamefont {Kim}, \citenamefont {Grätzel},\ and\
  \citenamefont {Maier}}]{acsenergylett.9b01605}%
  \BibitemOpen
  \bibfield  {author} {\bibinfo {author} {\bibfnamefont {A.}~\bibnamefont
  {Senocrate}}, \bibinfo {author} {\bibfnamefont {G.~Y.}\ \bibnamefont {Kim}},
  \bibinfo {author} {\bibfnamefont {M.}~\bibnamefont {Grätzel}}, \ and\
  \bibinfo {author} {\bibfnamefont {J.}~\bibnamefont {Maier}},\ }\href@noop {}
  {\bibfield  {journal} {\bibinfo  {journal} {ACS Energy Letters}\ }\textbf
  {\bibinfo {volume} {4}},\ \bibinfo {pages} {2859} (\bibinfo {year}
  {2019})}\BibitemShut {NoStop}%
\bibitem [{\citenamefont {Latini}\ \emph {et~al.}(2017)\citenamefont {Latini},
  \citenamefont {Gigli},\ and\ \citenamefont {Ciccioli}}]{C7SE00114B}%
  \BibitemOpen
  \bibfield  {author} {\bibinfo {author} {\bibfnamefont {A.}~\bibnamefont
  {Latini}}, \bibinfo {author} {\bibfnamefont {G.}~\bibnamefont {Gigli}}, \
  and\ \bibinfo {author} {\bibfnamefont {A.}~\bibnamefont {Ciccioli}},\
  }\href@noop {} {\bibfield  {journal} {\bibinfo  {journal} {Sustainable Energy
  Fuels}\ }\textbf {\bibinfo {volume} {1}},\ \bibinfo {pages} {1351} (\bibinfo
  {year} {2017})}\BibitemShut {NoStop}%
\bibitem [{\citenamefont {Besleaga}\ \emph {et~al.}(2016)\citenamefont
  {Besleaga}, \citenamefont {Abramiuc}, \citenamefont {Stancu}, \citenamefont
  {Tomulescu}, \citenamefont {Sima}, \citenamefont {Trinca}, \citenamefont
  {Plugaru}, \citenamefont {Pintilie}, \citenamefont {Nemnes}, \citenamefont
  {Iliescu}, \citenamefont {Svavarsson}, \citenamefont {Manolescu},\ and\
  \citenamefont {Pintilie}}]{doi:10.1021/acs.jpclett.6b02375}%
  \BibitemOpen
  \bibfield  {author} {\bibinfo {author} {\bibfnamefont {C.}~\bibnamefont
  {Besleaga}}, \bibinfo {author} {\bibfnamefont {L.~E.}\ \bibnamefont
  {Abramiuc}}, \bibinfo {author} {\bibfnamefont {V.}~\bibnamefont {Stancu}},
  \bibinfo {author} {\bibfnamefont {A.~G.}\ \bibnamefont {Tomulescu}}, \bibinfo
  {author} {\bibfnamefont {M.}~\bibnamefont {Sima}}, \bibinfo {author}
  {\bibfnamefont {L.}~\bibnamefont {Trinca}}, \bibinfo {author} {\bibfnamefont
  {N.}~\bibnamefont {Plugaru}}, \bibinfo {author} {\bibfnamefont
  {L.}~\bibnamefont {Pintilie}}, \bibinfo {author} {\bibfnamefont {G.~A.}\
  \bibnamefont {Nemnes}}, \bibinfo {author} {\bibfnamefont {M.}~\bibnamefont
  {Iliescu}}, \bibinfo {author} {\bibfnamefont {H.~G.}\ \bibnamefont
  {Svavarsson}}, \bibinfo {author} {\bibfnamefont {A.}~\bibnamefont
  {Manolescu}}, \ and\ \bibinfo {author} {\bibfnamefont {I.}~\bibnamefont
  {Pintilie}},\ }\href@noop {} {\bibfield  {journal} {\bibinfo  {journal} {The
  Journal of Physical Chemistry Letters}\ }\textbf {\bibinfo {volume} {7}},\
  \bibinfo {pages} {5168} (\bibinfo {year} {2016})}\BibitemShut {NoStop}%
\bibitem [{\citenamefont {Zhang}\ \emph {et~al.}(2019)\citenamefont {Zhang},
  \citenamefont {Liu}, \citenamefont {Xu}, \citenamefont {Ye}, \citenamefont
  {Li}, \citenamefont {Hu}, \citenamefont {Yang},\ and\ \citenamefont
  {Liu}}]{C8TC06129G}%
  \BibitemOpen
  \bibfield  {author} {\bibinfo {author} {\bibfnamefont {Y.}~\bibnamefont
  {Zhang}}, \bibinfo {author} {\bibfnamefont {Y.}~\bibnamefont {Liu}}, \bibinfo
  {author} {\bibfnamefont {Z.}~\bibnamefont {Xu}}, \bibinfo {author}
  {\bibfnamefont {H.}~\bibnamefont {Ye}}, \bibinfo {author} {\bibfnamefont
  {Q.}~\bibnamefont {Li}}, \bibinfo {author} {\bibfnamefont {M.}~\bibnamefont
  {Hu}}, \bibinfo {author} {\bibfnamefont {Z.}~\bibnamefont {Yang}}, \ and\
  \bibinfo {author} {\bibfnamefont {S.~F.}\ \bibnamefont {Liu}},\ }\href@noop
  {} {\bibfield  {journal} {\bibinfo  {journal} {J. Mater. Chem. C}\ }\textbf
  {\bibinfo {volume} {7}},\ \bibinfo {pages} {1584} (\bibinfo {year}
  {2019})}\BibitemShut {NoStop}%
\bibitem [{\citenamefont {Cai}\ \emph {et~al.}(2018)\citenamefont {Cai},
  \citenamefont {Wang}, \citenamefont {Seo},\ and\ \citenamefont
  {Yan}}]{doi:10.1063/1.5023797}%
  \BibitemOpen
  \bibfield  {author} {\bibinfo {author} {\bibfnamefont {P.}~\bibnamefont
  {Cai}}, \bibinfo {author} {\bibfnamefont {X.}~\bibnamefont {Wang}}, \bibinfo
  {author} {\bibfnamefont {H.~J.}\ \bibnamefont {Seo}}, \ and\ \bibinfo
  {author} {\bibfnamefont {X.}~\bibnamefont {Yan}},\ }\href@noop {} {\bibfield
  {journal} {\bibinfo  {journal} {Applied Physics Letters}\ }\textbf {\bibinfo
  {volume} {112}},\ \bibinfo {pages} {153901} (\bibinfo {year}
  {2018})}\BibitemShut {NoStop}%
\bibitem [{\citenamefont {Min}\ \emph {et~al.}(2020)\citenamefont {Min},
  \citenamefont {Hossain}, \citenamefont {Ma},\ and\ \citenamefont
  {Kaul}}]{min2020fabrication}%
  \BibitemOpen
  \bibfield  {author} {\bibinfo {author} {\bibfnamefont {M.}~\bibnamefont
  {Min}}, \bibinfo {author} {\bibfnamefont {R.~F.}\ \bibnamefont {Hossain}},
  \bibinfo {author} {\bibfnamefont {L.-C.}\ \bibnamefont {Ma}}, \ and\ \bibinfo
  {author} {\bibfnamefont {A.~B.}\ \bibnamefont {Kaul}},\ }\href@noop {}
  {\bibfield  {journal} {\bibinfo  {journal} {Journal of Vacuum Science \&
  Technology A: Vacuum, Surfaces, and Films}\ }\textbf {\bibinfo {volume}
  {38}},\ \bibinfo {pages} {052202} (\bibinfo {year} {2020})}\BibitemShut
  {NoStop}%
\bibitem [{\citenamefont {Chang}\ \emph {et~al.}(2018)\citenamefont {Chang},
  \citenamefont {Lin}, \citenamefont {Chen}, \citenamefont {Kuo},\ and\
  \citenamefont {Wang}}]{chang2018facile}%
  \BibitemOpen
  \bibfield  {author} {\bibinfo {author} {\bibfnamefont {Y.-H.}\ \bibnamefont
  {Chang}}, \bibinfo {author} {\bibfnamefont {J.-C.}\ \bibnamefont {Lin}},
  \bibinfo {author} {\bibfnamefont {Y.-C.}\ \bibnamefont {Chen}}, \bibinfo
  {author} {\bibfnamefont {T.-R.}\ \bibnamefont {Kuo}}, \ and\ \bibinfo
  {author} {\bibfnamefont {D.-Y.}\ \bibnamefont {Wang}},\ }\href@noop {}
  {\bibfield  {journal} {\bibinfo  {journal} {Nanoscale research letters}\
  }\textbf {\bibinfo {volume} {13}},\ \bibinfo {pages} {1} (\bibinfo {year}
  {2018})}\BibitemShut {NoStop}%
\end{thebibliography}%

\end{document}